\documentclass[lettersize,journal]{IEEEtran}
\newcommand{\CLASSINPUTbaselinestretch}{1.35}
\usepackage{cite}
\usepackage{amsmath,amssymb,amsfonts}
\usepackage{algorithm}
\usepackage{algpseudocode}
\usepackage{graphicx}
\usepackage{textcomp}
\usepackage{xcolor}
\usepackage{mathtools}
\usepackage{tikz}
\usetikzlibrary{shapes,arrows}
\usepackage{subcaption}
\usepackage{xcolor,soul}
\usepackage{hyperref}
\usepackage{cases}
\usepackage{siunitx}
\input{mysymbol.sty}
\newtheorem{problem}{\hspace{-10pt}\bf Problem}
\newtheorem{remark}{\hspace{-10pt}\textit{Remark}}

\def\BibTeX{{\rm B\kern-.05em{\sc i\kern-.025em b}\kern-.08em
    T\kern-.1667em\lower.7ex\hbox{E}\kern-.125emX}}

\begin{document}

\title{Annealed Langevin Dynamics for Massive MIMO Detection\\
\thanks{This work was partially supported by Nvidia. Email: \{nzilberstein, doost, ashu, segarra\}@rice.edu, cdick@nvidia.com. Preliminary results were published in~\cite{zilberstein2022}.}
}

\author{%
  \IEEEauthorblockN{Nicolas Zilberstein$^{\star}$, Chris Dick$^{\dagger}$, Rahman Doost-Mohammady$^{\star}$, Ashutosh Sabharwal$^{\star}$, Santiago Segarra$^{\star}$\\}
  \IEEEauthorblockA{$^{\star}$Rice University, USA \hspace{4cm}
                    $^{\dagger}$Nvidia, USA}  
}

\markboth{IEEE TRANSACTIONS ON WIRELESS COMMUNICATIONS (ACCEPTED)}%
{Zilberstein \MakeLowercase{\textit{et al.}}: \theTitle}

\maketitle
\begin{abstract}
Solving the optimal symbol detection problem in multiple-input multiple-output (MIMO) systems is known to be NP-hard.
Hence, the objective of any detector of practical relevance is to get reasonably close to the optimal solution while keeping the computational complexity in check.
In this work, we propose a MIMO detector based on an {annealed} version of Langevin (stochastic) dynamics.
More precisely, we define a stochastic dynamical process whose stationary distribution coincides with the posterior distribution of the symbols given our observations.
In essence, this allows us to approximate the maximum a posteriori estimator of the transmitted symbols by sampling from the proposed Langevin dynamic.
Furthermore, we carefully craft this stochastic dynamic by gradually adding a sequence of noise with decreasing variance to the trajectories, which ensures that the estimated symbols belong to a pre-specified discrete constellation.
Based on the proposed MIMO detector, we also design a robust version of the method by unfolding and parameterizing one term -- the score of the likelihood -- by a neural network.
Through numerical experiments in both synthetic and real-world data, we show that our proposed detector yields state-of-the-art symbol error rate performance and the robust version becomes noise-variance agnostic.
\end{abstract}
%

\section{Introduction}\label{S:intro}
Massive multiple-input multiple-output (MIMO) is a key technology for the development of modern and future wireless communications~\cite{mimoreview1}, \cite{mimoreview2}.
It is expected to play a key role in moving from the fifth to the sixth generation of cellular communications by achieving high data rates and spectral efficiency~\cite{6g}.
In massive MIMO systems, base stations are equipped with a large number of antennas, enabling them to handle several users simultaneously on the same time-frequency resource. 
Although this is a key enabler for achieving a great throughput and spectral efficiency, it comes at the cost of drastically increasing the complexity of the receiver.
Therefore, the design of an effective and low-complexity massive MIMO symbol detector is a major challenge, which is the focus of this paper.

In massive MIMO systems, the combination of high-order modulation schemes with a large-scale network -- multiple receivers and multiple users -- makes exact MIMO detection an intractable problem~\cite{Pia2017MixedintegerQP}. 
Indeed, given $N_u$ users and a modulation of $K$ symbols, the exact maximum likelihood (ML) estimator has an exponential decoding complexity $\mathcal{O}(K^{N_u})$.
Thus, computing the ML estimation is computationally prohibitive even for moderately-sized systems. 
Consequently, several approximate solutions for symbol detection have been proposed in the classical literature.
Two of the simplest detectors are zero forcing (ZF) and minimum mean squared error (MMSE)~\cite{Proakis2007}.
Although both (linear) detectors have low complexity and perform well on small systems, their performance degrades significantly for larger systems and higher modulation schemes~\cite{chockalingam_rajan_2014}.
Another classical detector is approximate message passing (AMP), which is asymptotically optimal for large MIMO systems with Gaussian channels; however, its performance degrades significantly for other practical channel distributions~\cite{amp}. 

In the past few years, several massive MIMO symbol detectors based on machine learning -- and, in particular, deep learning -- have been derived.
They have shown promising results, with a good balance between performance and complexity.
They can be roughly classified into data-driven methods, like DetNet~\cite{detNet2017}, and model-driven methods, like RE-MIMO~\cite{remimo}, OAMPNet~\cite{oampnet}, and hyperMIMO~\cite{zilberstein2021robust, hypermimo}.
The main issue with all of the learning-based detectors is their dependency on a particular configuration of the communication system as they are trained for a specific modulation scheme and, with the exception of RE-MIMO, for a given system size.


Given that obtaining the ML estimator is computationally prohibitive for large systems, an alternative approach for solving the MIMO detection problem is to generate samples from the true posterior distribution using Markov chain Monte Carlo (MCMC) methods~\cite[Chapter~8]{chockalingam_rajan_2014}.
A low-complexity MCMC family of samplers are the ones based on the Langevin dynamic~\cite{dalalyan2019, durmus_moulines}.
This iterative technique enables sampling from a given distribution by leveraging the availability of the score function (the gradient of the log-probability density function).
Under mild conditions~\cite{wellinglang}, the classical acceptance/rejection step in MCMC methods can be omitted. 
Recent developments in the context of generative modeling and denoising in image processing have shown promising results.
In~\cite{ermon2019, ermon2020}, an \textit{annealed} Langevin dynamic is used in the context of generative modeling for images.
Assuming an unknown distribution of the images, the authors propose to parameterize the score function as a neural network and use the {annealed} Langevin dynamic to sample from the underlying probability distribution.
In \cite{kawar2021snips}, the authors proposed to solve noisy image inverse problems by sampling from the posterior.

In this paper, we introduce the \emph{first massive MIMO detector base on {annealed} Langevin dynamics}.
This MCMC-based method requires no training, can be applied to any observed channel, and can handle different numbers of users and mixed modulation schemes.
A unique contribution of our proposed detector is to include the discrete nature of the constellation in the exploration of the search space, which is key given that {the} ML {computation entails} an integer optimization problem.
Moreover, we propose a robust version of the detector that does not require information of the noise variance.
We achieve this noise-variance agnostic behavior by parameterizing the score of the likelihood by a score network and leveraging the algorithm unfolding framework~\cite{eldar2021}.

\vspace{0.7mm}
\noindent
{\bf Contributions.}
The contributions of this paper are threefold:\\
1) We propose a novel detector based on \emph{annealed} Langevin dynamics, allowing us to include information of our discrete prior in the exploration of the posterior distribution.\\
2) We propose a robust version of the Langevin detector based on the algorithm unfolding framework, which does not require information about the noise distribution.\\
3) Through numerical experiments, we evaluate the behavior of our method for different hyperparameter settings and demonstrate that the proposed detector achieves a lower symbol error rate (SER) than baseline methods for massive MIMO systems, both in simulated and real-world scenarios.

\vspace{0.7mm}
\noindent
{\bf Paper outline.}
In Section~\ref{sec:background}, we present the system model and formulate the MIMO detection problem. 
Given this formulation of the problem, we give an overview of previous detectors.
In Section~\ref{sec:langevin}, we briefly introduce the Langevin dynamic and provide a detailed description of our proposed detector.
In Section~\ref{sec:unrolling}, we describe the proposed unfolded detector.
We illustrate results through numerical experimentation, demonstrating the superior performance of our detector in comparison with other benchmark methods in Section~\ref{sec:results}
Finally, Section~\ref{sec:conclusions} wraps-up the paper and discusses possible future work.



\section{Problem formulation and existing MIMO detectors}
\label{sec:background}
\vspace{2mm}
\subsection{System model and problem formulation}

We consider a communication channel with $N_u$ single-antenna transmitters or users and a receiving base station with $N_r$ antennas. 
The forward model for this MIMO system is given by
\begin{equation}\label{E:mimo_model_complex}
	\bar{\bby} = \bar{\bbH} \bar{\bbx} + \bar{\bbz},
\end{equation}
where $\bar{\bbH} \in \mathbb{C}^{N_r \times N_u}$ is the channel matrix, $\bar{\bbz} \sim \mathcal{CN}(\bb0, \sigma_0^2 \bbI_{N_r})$ is a vector of complex circular Gaussian noise, $\bar{\bbx} \in \mathcal{X}^{N_u}$ is the vector of transmitted symbols, $\mathcal{X}$ is a finite set of constellation points, and $\bby \in \mathbb{C}^{N_r}$ is the received vector.
Throughout this work we consider an equivalent real-valued representation obtained by considering the real $\mathfrak{R}(.)$ and imaginary $\mathfrak{I}(.)$ parts separately.
Denote $\bbx = [\mathfrak{R}(\bar{\bbx})^\top \mathfrak{I}(\bar{\bbx})^\top]$, $\bby = [\mathfrak{R}(\bar{\bby})^\top \mathfrak{I}(\bar{\bby})^\top]$, $\bbz = [\mathfrak{R}(\bar{\bbz})^\top \mathfrak{I}(\bar{\bbz})^\top]$ and 
\begin{equation}
\bbH = 
\begin{bmatrix}
    \mathfrak{R}(\bar{\bbH}) & -\mathfrak{I}(\bar{\bbH})\\
    \mathfrak{I}(\bar{\bbH}) & \mathfrak{R}(\bar{\bbH})\\
\end{bmatrix}.
\end{equation}
The system can be rewritten in the equivalent real-valued representation as follows\footnote{We define the noise as complex Gaussian circularly symmetric noise with variance $\sigma_0^2$, so each component has variance $\sigma_0^2/2$. However, we will consider throughout the work that $\sigma_0$ is the variance of each component as it is just a matter of re-scaling.}
\begin{equation}\label{E:mimo_model}
	\bby = \bbH \bbx + \bbz,
\end{equation}
In this work, a quadrature amplitude modulation (QAM) is used and symbols are normalized to attain unit average power. 
Every symbol is assumed to have the same probability of being chosen by the users.
Also, although we adopt a common constellation $\mathcal{X}$ for all users in our baseline problem formulation, we relax this requirement in our numerical experiments (see Section~\ref{sec:results}).
Moreover, perfect channel state information (CSI) is assumed, which means that $\bbH$ and $\sigma_0^2$ are known at the receiver, with the exception of Section~\ref{subsec:results_unf}.\footnote{To avoid notation overload, we adopt the convention that whenever we assume $\bbH$ to be known, $\sigma_0^2$ is also known.} 
Under this setting, the MIMO detection problem can be defined as follows.

\vspace{2mm}
\begin{problem}\label{P:main} \emph{
	Given perfect CSI and an observed $\bby$ following~\eqref{E:mimo_model}, find an estimate of $\bbx$.}
\end{problem}
\vspace{2mm}

Given the stochastic nature of $\bbz$ in~\eqref{E:mimo_model}, a natural way of solving Problem~\ref{P:main} is to search for the $\bbx$ that maximizes its \textit{posterior} probability given the observations $\bby$.
Hence, the optimal decision rule can be written as
\begin{align}\label{eq:map}
	\hat{\bbx}_{\mathrm{MAP}} &= \argmax_{\bbx \in \mathcal{X}^{N_u}}\,\, p(\bbx|\bby,\bbH)\\
	&= \argmax_{\bbx \in \mathcal{X}^{N_u}}\,\, p_{\bbz}(\bby - \bbH\bbx)p(\bbx)\nonumber,
\end{align}
%
\noindent where we applied the Bayes' rule. 
We assume that the symbols' prior distribution is uniform among the constellation elements and the measurement noise $\bbz$ is Gaussian.
As a result, the maximum a posteriori (MAP) detector reduces to an ML detector. 
Specifically, \eqref{eq:map} reduces to solving the following optimization problem
\begin{equation}\label{eq:ml}
	\hat{\bbx}_{\mathrm{ML}} = \argmin_{\bbx \in \mathcal{X}^{N_u}}\,\, ||\bby - \bbH\bbx||^2_2,
\end{equation}
%

\noindent which is NP-hard due to the finite constellation constraint $\bbx \in \mathcal{X}^{N_u}$, rendering $\hat{\bbx}_{\mathrm{ML}}$ intractable for practical applications.
Consequently, several schemes have been proposed in the past decades to provide efficient approximate solutions to Problem~\ref{P:main}, as mentioned in Section~\ref{S:intro} and further detailed in Section~\ref{Ss:related_work}.
In this paper, we propose to solve Problem~\ref{P:main} by (approximately) sampling from the posterior distribution in~\eqref{eq:map} using an annealed Langevin dynamic.

\subsection{Related work}
\label{Ss:related_work}
Many MIMO detection algorithms that approximate the optimal solution~\eqref{eq:ml} are iterative, where each iteration consists of the following two general steps
\begin{subequations}
\label{eq:general_iter}
\begin{align}\label{eq:general_iter_1}
    \bbu_t &= \hat{\bbx}_t + \bbA_t(\bby-\bbH\hat{\bbx}_t) + \bbb_t,\\
    \hat{\bbx}_{t+1} &= \eta_t(\bbu_t, \sigma_0^2)\label{eq:general_iter_2}.
\end{align}
\end{subequations}
The first $\textit{linear}$ step \eqref{eq:general_iter_1} takes as input the tuple $(\bby, \bbH, \hat{\bbx}_t)$ and applies a linear transformation $\bbA_t$ to obtain an intermediate signal $\bbu_t$ in a continuous domain.
The second $\textit{non-linear}$ step \eqref{eq:general_iter_2} applies a \textit{denoiser}  $\eta_t$ to the intermediate signal to get the estimation $\hat{\bbx}_{t+1}$ of the next iteration. 
Hence, the estimation is composed by an optimization in a continuous domain followed by a (possibly approximate) projection to the discrete constellation set.

A common choice for the denoiser is the minimizer of $\mathbb{E}[||\eta_t(\bbu_t, \sigma_0^2) - \bbx||_2|\bbu_t]$, given by the conditional expectation $ \eta_t(\bbu_t, \sigma_0^2) = \mathbb{E}[\bbx|\bbu_t]$.
For the i.i.d. Gaussian noise case, it boils down to an element-wise function with the following expression

\begin{align}\label{eq:denoiser}
	\mathbb{E}[x_j|[\bbu_t]_j] &= \frac{1}{Z}\sum_{x_k \in \ccalX} x_k \exp\bigg(\frac{-||[\bbu_t]_j - x_k||^2}{2\sigma_0^2}\bigg),
\end{align}

\noindent where $Z = \sum_{x_k \in \ccalX} \exp\big(\frac{-||[\bbu_t]_j - x_k||^2}{2\sigma_0^2}\big)$.
Notice that~\eqref{eq:denoiser} depends on $\bbu_t$ and the \textit{noise variance} $\sigma_0$. 
We now describe several detectors, both classical and learning-based, and we show that many of these approximate solutions represent particular cases of~\eqref{eq:general_iter}.

\vspace{3mm}
\noindent{\bf Classical MIMO detectors.} The classical body of work (ZF, MMSE, AMP) can be described as particular cases of the framework in~\eqref{eq:general_iter}.
For instance, ZF is equivalent to a single-step in~\eqref{eq:general_iter_1}, where $\bbA_0 = \bbH^{\dagger}, \bbb_0 = 0$ and $\hat{\bbx}_0 = 0$.
MMSE detector, which corresponds to regularized version of ZF, is described by setting $\bbA_0 = (\bbH^\top\bbH + \sigma_0^2\bbI_{2N_u})^{-1}, \bbb_0 = 0$ and $\hat{\bbx}_0 = 0$.
Both methods solve first a relaxation of the original problem, by defining $\bbx \in \ccalX^{N_u}$ as $\bbx \in \mathbb{C}^{N_u}$. 
After this, the final solution is obtained by rounding the intermediate estimation onto one of the constellation elements.
Although both are practical because of their low complexity, their performance degrades in massive MIMO systems.

A more sophisticated classical method is AMP, which can also be described as a particular case of~\eqref{eq:general_iter}, where $\bbA_t = \bbH^{H}$ and $\bbb_t$ is known as the Onsager term~\cite{amp2}. 
This method represents the MIMO system as a graphical model and performs approximate inference instead of belief propagation as it is prohibitive for massive MIMO systems.
It presents a good balance between complexity and performance, as it is asymptotically optimal with i.i.d. Gaussian channels and it has low complexity.
However, its behavior depends on the channel model, so its performance degrades significantly for other channel matrices that are not i.i.d. Gaussian.
As a consequence, a generalization of AMP, known as Orthogonal AMP (OAMP), was proposed in~\cite{oamp}.
It relaxes the constraint on channel matrices imposed by AMP to unitary-invariant matrices~\cite{unitaryMatrix}. 
Other low-complexity detectors are V-BLAST~\cite{V-blast} and Semi-Definite Relaxation (SDR)~\cite{SDR1}. 
However, neither of the methods balances low complexity and good performance to be on par with the optimal detector (ML).

While classical methods seek to solve the optimization problem, an alternative solution to Problem~\ref{P:main} is to generate samples of the posterior distribution.
This can be done using MCMC methods~\cite[Chapter~8]{chockalingam_rajan_2014}. 
For instance, a detector based on the Gibbs Sampler was presented in~\cite{MCMC_gibbs}.
Recently, in~\cite{MIMOmetropolis}, a detector based on Metropolis-Hastings was proposed. 
The idea of this detector is to explore the search space following a random walk in the gradient descent direction. 
Notice that this method can be described by~\eqref{eq:general_iter} by adding a randomized step to generate random samples.
Lastly, a detector using the Langevin dynamics was independently proposed in~\cite{langevinMIMO} (concurrently made available with our preliminary work~\cite{zilberstein2022}), but the approach is significantly different from the one presented in this paper, as they do not adopt an annealing process to incorporate the discrete nature of the problem, they are constrained to a particular modulation scheme, and they do not propose a robust learning-based version rooted in algorithm unfolding.

\vspace{3mm}
\noindent{\bf Learning-based MIMO detectors.} The application of deep learning to MIMO detection is a body of work that has gained traction in the last few years. 
The basic idea is to learn a map -- a function approximator -- from the space of observations $\bby$ and CSI $\bbH$ to the corresponding (approximate) transmitted symbols $\bbx$. 
In this way, when a new observation $\bby$ is received (along with the CSI), $\bbx$ can be efficiently estimated using the learned map.
Based on how the learned map is designed, one can roughly categorize each detector as either data-driven or model-driven detector.


MMNet~\cite{mmnet} is a low complexity channel-specific detector based on the iterative soft-thresholding algorithm~\cite{ISTA}.
It can be described as in~\eqref{eq:general_iter} by replacing $\bbA_t$ with a matrix of learning parameters, setting $\bbb_t = 0$, and adding a vector of learning parameters to the noise estimation in the denoiser~\eqref{eq:denoiser}.
Although its performance is comparable to the ML for i.i.d. Gaussian channels, for correlated ones the detector needs to be trained online for each channel realization.
Hence, it is unsuitable for real-time applications.
However, this issue can be alleviated by using HyperNetworks~\cite{hypermimo, zilberstein2021robust}.

DetNet, proposed in ~\cite{detNet2017}, is a data-driven detector inspired by iterative projected gradient descent. 
Its performance is very good on i.i.d. Gaussian channels with low-order modulation schemes, but it degrades for correlated channels and higher-order modulations. 

OAMPNet~\cite{oampnet} is a model-driven detector based on the OAMP algorithm.
It adds two learning parameters per iteration of the OAMP algorithm.
For i.i.d. Gaussian channels, its performance is similar to the ML.
However, it performs poorly in more realistic channels.
Similar to MMNet, it can be described by~\eqref{eq:general_iter}.

RE-MIMO~\cite{remimo} is a detector inspired by the Recurrence Inference Machine framework\cite{rim} that achieves state-of-the-art performance for correlated channel models. 
It is an iterative detector with two steps.
Each iteration is composed by an encoder step, parameterized by a transformer, and a predictor module, which plays the role of the denoiser. 
In addition to its good performance, the detector can handle a varying number of transmitters using the same trained model.
However, the number of parameters is high, and from the practical point of view, it does not handle different modulations simultaneously, and its performance degrades severally under noise uncertainty.

\vspace{1mm}
Overall, both classical and learning-based detectors follow an iterative approach that splits the problem into two steps, where the optimization is done in a continuous space followed by a projection step.
Hence, we are motivated by the following question: \textit{Can we replace the two-step iterative framework in~\eqref{eq:general_iter} with a single-step (per iteration) that still leverages the prior information about the discrete constraint?}
We achieve this objective by leveraging Langevin dynamics.

In essence, we seek a detector that solves Problem~\ref{P:main} directly by efficiently searching in the discrete search space.
At a high level, we propose to approximate the distribution of the symbols by a sequence of Gaussian noise with mean at each symbol and decreasing variance, thus concentrating around delta distributions at each symbol.
Our main contribution is the use of the above approximation procedure  to \textit{define the gradient of the log-posterior} and to leverage the \textit{Langevin dynamics} -- a low complexity MCMC based sampler to generate samples from the posterior (presented in Section~\ref{sec:langevin}).
After designing a detector with the above characteristics, the next natural question is: \textit{Is our proposed detector robust against noise uncertainty? If it is not, can we design a detector that still performs well under noise uncertainty?}
To answer this question, we modify our design by adding a learning process based on the unfolding framework, which is our second contribution and is presented in Section~\ref{sec:unrolling}.

\section{Langevin for MIMO detection}
\label{sec:langevin}

In Section~\ref{subsec:langevindyn} we briefly introduce the Langevin dynamic while in Section~\ref{subsec:posterior} we explain how we propose to use it for MIMO detection.
In particular, we detail the expressions of the score functions involved in the sampling process to solve Problem~\ref{P:main}.

\subsection{Langevin dynamics}
\label{subsec:langevindyn}

The Langevin diffusion is the Markov process on $\mathbb{R}^d$ that solves the stochastic differential equation
\begin{align}\label{eq:ct_underdamped_langevin}
    \text{d}\bbx_t &= -\nabla U(\bbx_t) + \sqrt{2\tau}\text{d}\bbW,
\end{align}
where $\bbW$ is a standard $d$-dimensional Brownian motion, $U \in \ccalC^2(\mathbb{R}^d)$ is called the potential, and $\tau$ is a temperature parameter.
The dynamic in~\eqref{eq:ct_underdamped_langevin} describes the dynamic of a particle that moves according to Newton's second law, where the force is given by $\nabla U(\bbx_t)$ and is in contact with a thermal reservoir, i.e., collisions with other particles, represented by the random term (Brownian motion).
Under mild conditions, it can be shown that the invariant distribution $\pi(\bbx)$ of the continuous-time process is proportional to $\exp (-\tau^{-1}U(\bbx))$~\cite{pavliotis_book}.
In particular, given some target distribution $p(\bbx)$ from which we want to generate samples $\bbx \in \mathbb{R}^d$ and fixing $\tau = 1$, if we define $U(\bbx) = -\log p(\bbx)$, then~\eqref{eq:ct_underdamped_langevin} defines a MCMC sampler as $\pi(x) \propto p(\bbx)$.
This entails an MCMC algorithm \cite{MCMCbook, Roberts1996ExponentialCO} known as the unadjusted Langevin algorithm, which is obtained as the Euler-Maruyama discretization of~\eqref{eq:ct_underdamped_langevin} and is described by the following discrete-time equation

\begin{equation}\label{eq:langevin}
	\bbx_{t+1} = \bbx_t + \epsilon \nabla_{\bbx_t}\log p(\bbx_t) + \sqrt{2\epsilon}\, \bbw_t,
\end{equation}
where $p(\bbx)$ is the target distribution from which we want to generate samples $\bbx \in \mathbb{R}^N$ and $\bbw_t \sim \ccalN(0, \bbI_N)$. 
The dynamic in~\eqref{eq:langevin} explores the domain of the target distribution by moving in the direction of the gradient of the logarithm of the target density $\nabla_{\bbx}\log p(\bbx)$, known as the \textit{score function}.
It is essentially a combination of a stochastic gradient ascent in the direction of the score function with a noise term, which prevents the method from collapsing to a local maxima.
Under some regularity conditions~\cite{wellinglang}, the distribution of $\bbx_T$ converges to $p(\bbx)$ when $\epsilon \rightarrow 0$ and $T \rightarrow \infty$, in which case $\bbx_T$ becomes an exact sample from $p(\bbx)$. 
In practice, neither $\epsilon \rightarrow 0$ nor $T \rightarrow \infty$, so a Metropolis-Hastings acceptance/rejection step is used to ensure convergence, leading to the so-called Metropolis-adjusted Langevin algorithm (MALA)~\cite{Roberts1996ExponentialCO}.
An alternative, proposed in~\cite{wellinglang}, is a time-inhomogeneous variant of~\eqref{eq:langevin}, i.e., defining a variable step size $\epsilon_t$.
Thus, when $\epsilon_t$ decreases to zero for large $t$, the error becomes negligible and the acceptance/rejection step can be omitted.
Although this result is asymptotic, also non-asymptotic convergence results in the 2-Wasserstein distance have been obtained when the target distribution is log-concave and smooth (the potential has Lipschitz continuous gradient and is strongly convex)~\cite{dalalyan2019}.
It should be noted that the \textit{only} requirement for sampling from $p(\bbx)$ using this procedure is knowing the score function.

\vspace{2mm}
\noindent {\bf Temperature parameter.} The discretization of the Langevin dynamic in~\eqref{eq:ct_underdamped_langevin} can be rewritten for a temperature parameter $\tau \neq 1$, that adds a new degree of freedom:

\begin{equation}
    \bbx_{t+1} = \bbx_t + \epsilon \nabla_{\bbx_t}\log p(\bbx_t) + \sqrt{2\epsilon \tau}\, \bbw_t.
\end{equation}

\noindent The behaviour of this dynamic is the same as in~\eqref{eq:langevin}, but the stationary distribution is now $p(\bbx)^{1/\tau}$; see~\cite[Chapter~6]{pavliotis_book} for details.
This additional degree of freedom provides a trade-off between mixing time and fidelity.
Indeed, even though the stationary distribution is modified, when $\tau \neq 1$ the mixing time of the dynamic can be improved.
This is especially helpful for multimodal distributions.

\subsection{Detection by sampling from the posterior distribution}
\label{subsec:posterior}

Problem~\ref{P:main} is intractable due to the finite constellation constraint.
Therefore, instead of solving the problem using an approximate solution like the ones described in Section~\ref{Ss:related_work}, we propose to \emph{generate a set of samples that approximately come from the posterior distribution $p(\bbx|\bby, \bbH)$ using~\eqref{eq:langevin} and then select the one that minimizes the objective in~\eqref{eq:ml}}.
The key ingredient in the Langevin dynamic is the score function, which for our case is given by $\nabla_{\bbx}\log p(\bbx| \bby, \bbH)$.
After applying Bayes' rule, this score function can be rewritten as 
\begin{equation}\label{E:score_function}
\nabla_{\bbx}\log p(\bbx|\bby, \bbH) = \nabla_{\bbx}\log p(\bby|\bbx,\bbH) + \nabla_{\bbx}\log p(\bbx),
\end{equation}
where the term $\nabla_{\bbx}\log p(\bby|\bbx,\bbH)$ corresponds to the score function of the likelihood and $\nabla_{\bbx}\log p(\bbx)$ to the score function of the prior.
Notice that this latter term is not well defined due to the discrete nature of the symbols.

To circumvent this obstacle, inspired by~\cite{kawar2021snips}, we approximate the prior by using an \textit{annealed} version of the Langevin dynamic. 
Specifically, instead of working with the discrete symbols $\bbx$, we define a perturbed version of the symbols $\tilde{\bbx} = \bbx + \bbn$ with $\bbn \sim \mathcal{N}(0, \sigma^2\bbI)$, for different values of $\sigma^2$. 
Note that $\tilde{\bbx}$ is now continuous, so it is a random variable described by a probability density with a well-defined gradient.
This allows us to run a Langevin dynamic in $\tilde{\bbx}$ instead of $\bbx$.
Moreover, if we make $\sigma^2 \to 0$ then $\tilde{\bbx}$ concentrates around $\bbx$, allowing us to effectively sample from $p(\bbx|\bby, \bbH)$, as wanted.

In a nutshell, the algorithm works as follows. 
First, we initialize $\tilde{\bbx}_0$ uniformly at random in $[-1,1] \times [-1,1]$, since the symbols are assumed to be normalized. 
Then, we follow the direction of the score function of the log-posterior density of the perturbed symbol $\nabla_{\tilde{\bbx}}\log p(\tilde{\bbx}| \bby, \bbH)$, starting with a high $\sigma$ and gradually decreasing its value until $\tilde{\bbx} \approx \bbx$.
Apart from enabling the approximation of the score function of the prior distribution, the annealing process also improves the mixing time of the Langevin dynamic~\cite{ermon2019}.
This is particularly important in multimodal distributions.
Having introduced the high-level idea of our method, we now provide more details on the \emph{annealing process}, exact expressions for the terms in the \emph{score function}, and a step-by-step description of the \emph{algorithm}.

\vspace{1mm}

\begin{figure}[t]
	\centering
	\includegraphics[width=0.4\textwidth]{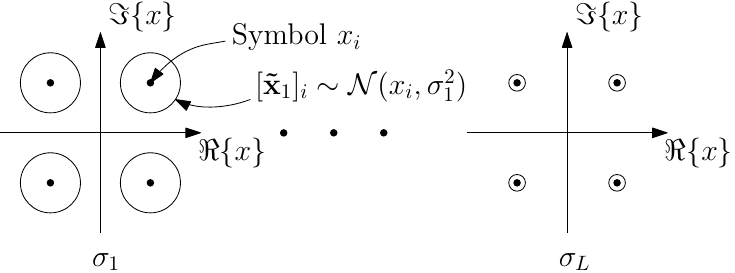}
	\vspace{-0.03in}
	\caption{{\small Scheme of the annealed process. We consider a QPSK constellation and at each level we add Gaussian noise. The variance of the noise decreases at higher levels. In the last level $L$, the Gaussian is very sharp around each symbol, mimicking our true discrete prior over the constellation.}}
	\vspace{-0.1in}
	\label{fig:gaussian_levels}
\end{figure}

\noindent{\bf Annealing process.}
We define a sequence of noise levels $\{\sigma_l\}_{l=1}^{L+1}$ such that $\sigma_1 > \sigma_2 > \cdots > \sigma_L > \sigma_{L+1} = 0$.
Then, at each level we define a perturbed version of the true symbols $\bbx$
\begin{equation}\label{eq:pert_symbs}
    \tilde{\bbx}_{l} = \bbx + \bbn_{l},
\end{equation}
where $\bbn_{l} \sim \mathcal{N}(0, \sigma_l^2\bbI)$. 
A representation (for a QPSK modulation) of this process is shown in Fig. \ref{fig:gaussian_levels}.
From the figure, we observe that the perturbed symbols defined in~\eqref{eq:pert_symbs} are defined in a continuous domain due the noise addition, and centered at each symbol.
Since the variance of the noise injected at each level is a predefined parameter, we design the sequence in such a way that the noise injected in the first levels is high enough to explore the search space, while at the last levels is very small, approximating the true discrete distribution given by a set of delta functions at each symbol with uniform weight.
Notice that we defined the last variance $\sigma_{L+1}$ as equal to $0$, although in practice it is never reached. 
However, it is necessary for the construction of the synthetic noise, which is explained next.

Given the perturbed symbols in~\eqref{eq:pert_symbs}, the forward model in~\eqref{E:mimo_model} can be rewritten as
\begin{align}\label{eq:forwardmodel_noise}
	\bby &= \bbH\tilde{\bbx}_l + (\bbz - \bbH\bbn_l).
\end{align}

\noindent In the new forward model in~\eqref{eq:forwardmodel_noise}, the likelihood is given by $p(\bby|\tilde{\bbx}_l, \bbH) = p(\bbz - \bbH {\bbn}_l|\tilde{\bbx}_l)$, which is not Gaussian: although $p(\bbn_l)$ is a Gaussian distribution, after conditioning on $\tilde{\bbx}_l$ the conditional distribution $p(\bbn_l|\tilde{\bbx}_l)$ is no longer Gaussian. 
To understand this, notice that~\eqref{eq:pert_symbs}, $p(\bbn_l|\tilde{\bbx}_l) \propto p(\tilde{\bbx}_l|\bbn_l)p(\bbn_l)$ is a mixture of Gaussians weighted by the distribution of $\bbx$.

However, an analytical expression for the score of the likelihood can still be obtained by following the approach in~\cite{kawar2021snips}, where a synthetic annealed noise, that is carved from the measurement noise $\bbz$, is constructed gradually.
Hence, before giving an expression of the score functions, we explain how the synthetic noise is defined.
Before moving to this explanation, and in order to get a tractable expression, we have to rely on the singular value decomposition (SVD) of the channel matrix given by $\bbH = \bbU\boldsymbol{\Sigma}\bbV^{\top}$ as well as in the spectral representation of $\tilde{\bbx}_l$, $\bby$, $\bbz$ and $\bbn_l$ defined as $\tilde{\boldsymbol{\chi}}_l = \bbV^{\top} \tilde{\bbx}_l$, $\boldsymbol{\eta} = \bbU^{\top}\bby$, $\boldsymbol{\zeta} = \bbU^{\top}\bbz$ and $\boldsymbol{\nu}_l = \bbV^{\top}\bbn_l$.
Given the spectral representation, the model in~\eqref{eq:forwardmodel_noise} can be rewritten as
\begin{align}\label{eq:forwardmodel_noise_spectral}
	\boldsymbol{\eta} &= \boldsymbol{\Sigma}\tilde{\boldsymbol{\chi}}_l + (\boldsymbol{\zeta} - \boldsymbol{\Sigma}\boldsymbol{\nu}_l).
\end{align}

The model defined in~\eqref{eq:forwardmodel_noise_spectral} is equivalent to~\eqref{eq:forwardmodel_noise} but in the SVD (spectral) domain.
Notice that expressing the system in the spectral domain as in~\eqref{eq:forwardmodel_noise_spectral} maintains the distributions involved invariant because $\bbU$ and $\bbV$ are orthogonal matrices.

\vspace{1mm}
\noindent {\bf Synthetic noise construction.}
The key idea of the synthetic noise is that it is carved from $\bbz$ gradually. 
To see what we formally mean by this, we assume that for every $j$ such that $s_j \neq 0$, with $s_j$ a singular value of $\bbH$, there exists an $l_j$ such that $\sigma_{l_j} s_j < \sigma_0$ and $\sigma_{l_j - 1} s_j > \sigma_0$.
Then, we define $\tilde{\bbx}_{L+1} = \bbx$ and for every $l=L,L-1,\cdots, 1$, $\tilde{\bbx}_l = \tilde{\bbx}_{l+1} + \boldsymbol{\omega}_l$ where $\boldsymbol{\omega}_l \sim \ccalN(0, (\sigma_l^2 - \sigma_{l+1}^2)\bbI)$.
Notice that with this definition, we recover the definition in~\eqref{eq:pert_symbs} by replacing recursively $\tilde{\bbx}_{l+1}$ and where $\bbn_l = \sum_{k=l}^L\boldsymbol{\omega}_k$.
Thus, the noise in~\eqref{eq:forwardmodel_noise_spectral} can be rewritten as
\begin{align}\label{eq:noise_spectral}
    \boldsymbol{\zeta} - \boldsymbol{\Sigma}\boldsymbol{\nu}_l  &= \boldsymbol{\zeta} - \boldsymbol{\Sigma}\bigg(\sum_{k=l}^L\bbV^{\top}\boldsymbol{\omega}_k\bigg),
\end{align}
Given~\eqref{eq:noise_spectral}, it follows that the annealed noise $\boldsymbol{\nu}_l$ is independent of $\tilde{\bbx}_l$; for more details, see~\cite[Appendix~A]{kawar2021snips}.
Moreover, given that both $\bbzeta$ and $\bbV^{\top}\boldsymbol{\omega}_k$ are Gaussian, the expression in~\eqref{eq:noise_spectral} is also Gaussian with $0$ mean and a covariance matrix given by
%
\begin{align}\label{eq:noise_covariance}
\mathbb{E}[(\boldsymbol{\zeta} - \boldsymbol{\Sigma}\boldsymbol{\nu}_l)(\boldsymbol{\zeta} - \boldsymbol{\Sigma}\boldsymbol{\nu}_l)^{\top}] = \mathbb{E}[\boldsymbol{\zeta}\boldsymbol{\zeta}^{\top}] +& \boldsymbol{\Sigma}\mathbb{E}[\boldsymbol{\nu}_l\boldsymbol{\nu}_l^{\top}]\boldsymbol{\Sigma}^{\top} \\ &- 2\mathbb{E}[\boldsymbol{\Sigma}\boldsymbol{\nu}_l\boldsymbol{\zeta}^{\top}]\nonumber.
\end{align}
The first two terms are known -- they are the covariance matrices of the measurement noise and the noise defined in~\eqref{eq:pert_symbs} --, and the third term can be rewritten as $\mathbb{E}[\boldsymbol{\Sigma}\boldsymbol{\nu}_l\boldsymbol{\zeta}^{\top}] = \mathbb{E}[\boldsymbol{\Sigma}(\sum_{k=l}^L\bbV^{\top}\boldsymbol{\omega}_k)\boldsymbol{\zeta}^{\top}] = \sum_{k=l}^L\mathbb{E}[\boldsymbol{\Sigma}\bbV^{\top}\boldsymbol{\omega}_k\boldsymbol{\zeta}^{\top}]$.

Hence, we need to define the statistical relationship between $\boldsymbol{\zeta}$ and $\boldsymbol{\Sigma}\bbV^{\top}\boldsymbol{\omega}_k$, which is the key step in this construction.
Essentially, we seek for an uncorrelated multivariate Gaussian distribution where the variance depends only on the relation between $\sigma_l s_j$ and $\sigma_0$, i.e., each direction in the spectral domain is independent of each other.
Hence, following~\cite{kawar2021snips}, we define the noise as (we focus on a single entry $j$)
\begin{align}\label{eq:cross_noise}
[\mathbb{E}[\boldsymbol{\Sigma}\bbV^{\top}\boldsymbol{\omega}_l\boldsymbol{\zeta}^{\top}]]_j = \\
 & 
   \hspace{-10mm} 
   \begin{cases}
   [\mathbb{E}[(\boldsymbol{\Sigma}\bbV^{\top}\boldsymbol{\omega}_l)(\boldsymbol{\Sigma}\bbV^{\top}\boldsymbol{\omega}_l)^\top]]_j, \hspace{7mm} l \geq l_j
  \\
  [\mathbb{E}[(\boldsymbol{\zeta} - \boldsymbol{\Sigma}\boldsymbol{\nu}_{l_j})(\boldsymbol{\zeta} - \boldsymbol{\Sigma}\boldsymbol{\nu}_{l_j})^\top]]_j, \,\,\,\, l = l_{j}-1 \\
  0, \hspace{45mm} l < l_j - 1.
\end{cases}
\nonumber
\end{align}
It can be checked by replacing~\eqref{eq:cross_noise} in~\eqref{eq:noise_covariance} that the distribution of $\boldsymbol{\zeta} - \boldsymbol{\Sigma}\boldsymbol{\nu}_l$ is multivariate Gaussian, where each component is distributed as $[\boldsymbol{\zeta} - \boldsymbol{\Sigma}\boldsymbol{\nu}_l]_j \sim \ccalN(0, |\sigma_0^2 - \sigma_l^2s_j^2|)$ for $j=1,\cdots, N_u$.
Intuitively, when $\sigma_i s_j < \sigma_0$, the synthetic noise is fully immersed in the measurement noise, while if $\sigma_i s_j > \sigma_0$ there is an independent portion that is added.

\vspace{1mm}
\noindent {\bf Score function.} With the spectral representation in mind, our goal is to run a Langevin dynamic whose score function for every noise level $l$ is given by [cf.~\eqref{E:score_function}]
\begin{equation}\label{E:score_function_spectral}
\nabla_{\tilde{\boldsymbol{\chi}}_l}\!\log p(\tilde{\boldsymbol{\chi}}_l| \boldsymbol{\eta}, \bbH) = \nabla_{\tilde{\boldsymbol{\chi}}_l} \log p(\boldsymbol{\eta}|\tilde{\boldsymbol{\chi}}_l, \bbH) + \nabla_{\tilde{\boldsymbol{\chi}}_l}\log p(\tilde{\boldsymbol{\chi}}_l).
\end{equation}
We now provide a closed-form expression for both constituent terms in this score function.
\vspace{1mm}

\noindent \emph{i) Score of the likelihood:} Based on the above discussion, the final expression for the score of the likelihood in the spectral domain is given by
\begin{equation}\label{eq:score_likeli}
    \nabla_{\tilde{\boldsymbol{\chi}}_l} \! \log p(\boldsymbol{\eta}|\tilde{\boldsymbol{\chi}}_l, \bbH)  = \boldsymbol{\Sigma}^\top \,\, |\sigma_0^2\bbI - \sigma_l^2\boldsymbol{\Sigma}\boldsymbol{\Sigma}^\top|^{\dagger}\,\, ( \boldsymbol{\eta} - \boldsymbol{\Sigma} \tilde{\boldsymbol{\chi}}_l).
\end{equation}
To give some intuition, the score function of the likelihood is given by the gradient of a multivariate Gaussian distribution: the residual error $( \boldsymbol{\eta} - \boldsymbol{\Sigma} \tilde{\boldsymbol{\chi}}_l) = (\bbU^\top\bby - \boldsymbol{\Sigma}\bbV^\top \tilde{\bbx}_l)$ is multiplied by the (pseudo-)inverse of the covariance matrix, which is a diagonal matrix with entries given by $|\sigma_0^2 - \sigma_l^2s_j^2|$. 

\vspace{1mm}
\noindent \emph{ii) Score of the annealed prior:} We first notice that $\nabla_{\tilde{\boldsymbol{\chi}}_l}\log p(\tilde{\boldsymbol{\chi}}_l) = \bbV^{\top}\nabla_{\tilde{\bbx}_l}\log p(\tilde{\bbx}_l)$ due to the orthogonality of $\bbV$.
Moreover, based on the Tweedie's identity~\cite{TweedieIdent}, we can relate the score function $\nabla_{\tilde{\bbx}_l}\log p(\tilde{\bbx}_l)$ and the MMSE denoiser as follows
\begin{equation}\label{eq:prior}
	\nabla_{\tilde{\bbx}_l}\log p(\tilde{\bbx}_l) = \frac{\mathbb{E}_{\sigma_l}[\bbx|\tilde{\bbx}_l] - \tilde{\bbx}_l}{\sigma_l^2}.
\end{equation}
In particular, the conditional expectation, defined as $\mathbb{E}_{\sigma_l}[\bbx|\tilde{\bbx}_l] = \sum_{\bbx_k \in \ccalX^{N_u}}\bbx_k p(\bbx_k|\tilde{\bbx}_l)$ \linebreak$= \frac{1}{p(\tilde{\bbx}_l)}\sum_{\bbx_k \in \ccalX^{N_u}} \bbx_k p(\tilde{\bbx}_l|\bbx_k)p(\bbx_k)$, can be calculated element-wise as
\begin{align}\label{E:mixed_gaussian}
	\mathbb{E}_{\sigma_l}[x_j|[\tilde{\bbx}_l]_j] &= \frac{1}{Z}\sum_{x_k \in \ccalX} x_k \exp\bigg(\frac{-||[\tilde{\bbx}_l]_j - x_k||^2}{2\sigma_l^2}\bigg),
\end{align}
where $Z = \sum_{x_k \in \ccalX} \exp\Big(\frac{-||[\tilde{\bbx}_l]_j - x_k||^2}{2\sigma_l^2}\Big)$ and $j=1,\cdots, N_u$. 

\vspace{2mm}
\noindent {\bf Algorithm.} 
Alg.~\ref{alg} illustrates how to generate samples $\hat{\bbx}$ from the (approximate) posterior $p(\bbx|\bby,\bbH)$.
Notice that we write the Langevin equation with the temperature parameter $\tau$, as it is more general; when $\tau = 1$, the dynamic boils down to the annealed Langevin dynamic in~\eqref{eq:langevin}.
As discussed in~\cite{kawar2021snips}, when computing the entries of the score function in~\eqref{E:score_function_spectral} using the expressions in~\eqref{eq:score_likeli} and~\eqref{eq:prior}, one of these terms might be negligible with respect to the other depending on the noise level.
Thus, the elementwise score of the posterior will be given by
 \begin{align}\label{eq:full_score}
 [\nabla_{\tilde{\boldsymbol{\chi}}_{l}}\!\log p(\tilde{\boldsymbol{\chi}}_{l}| \boldsymbol{\eta}, \bbH)]_j= \\
 & 
 \hspace{-3cm}
    \begin{cases}
  [\nabla_{\tilde{\boldsymbol{\chi}}_l} \! \log p(\boldsymbol{\eta}|\tilde{\boldsymbol{\chi}}_l, \bbH) + \bbV^{\top}\nabla_{\tilde{\bbx}}\log p(\tilde{\bbx})]_j, \,\,\,\, \sigma_0 \geq \sigma_ls_j
  \\
  [\nabla_{\tilde{\boldsymbol{\chi}}_l} \! \log p(\boldsymbol{\eta}|\tilde{\boldsymbol{\chi}}_l, \bbH)]_j,  \hspace{29mm} \sigma_0 < \sigma_ls_j \\
  [\bbV^{\top}\nabla_{\tilde{\bbx}} \! \log p(\tilde{\bbx})]_j, \hspace{34mm}  s_j = 0.
\end{cases}
\nonumber
\end{align}
Intuitively, when the injected noise at level $l$ is such that $\sigma_l s_j > \sigma_0$, then the contribution of the score of the prior is negligible and can be ignored, meaning that the Langevin dynamic considers \textit{the observation as a denoised estimation}.
Then, when $\sigma_l s_j < \sigma_0$, the estimation needs to be refined by \textit{considering prior information combined with the likelihood term}.
Similarly, whenever $s_j=0$, the corresponding entry $\eta_j$ is uninformative and the score of the likelihood can be ignored.
Furthermore, a refinement that we incorporate in the algorithm is the use of user-dependent step sizes.
Instead of using a constant \emph{scalar} step size $\epsilon$ as in~\eqref{eq:langevin} or even time-varying versions of it, in Alg.~\ref{alg} we employ user-dependent \emph{diagonal matrices} $\boldsymbol{\Lambda}_l$.
In this way, different entries of our vector-valued Langevin dynamic can be updated at different rates depending on the singular values of the channel under consideration.
Finally, given that $\sigma_L \neq 0$, the sample will be very close to the constellation but not exactly. 
Hence, we project our sample to the constellation by taking $\bar{\bbx} = \argmin_{\bbx \in \ccalX^{N_u}}||\bbx - \bbV\tilde{\boldsymbol{\chi}}_{T,L}||_2^2$.

Given that Alg.~\ref{alg} is stochastic, one can run the algorithm multiple times to generate several samples $\bar{\bbx}$ from the same (approximate) posterior distribution.
Hence, as we want to approximate the MAP estimate -- equivalently for this case, the ML estimate -- we run $M$ different Langevin trajectories for each pair $\{\bby, \bbH\}$ and keep the sample that minimizes~\eqref{eq:ml}.
Formally, given $M$ samples $\{\bar{\bbx}_m\}_{m=1}^M$ obtained from Alg.~\ref{alg}, our final estimate is given by
\begin{equation}\label{eq:Ntraj}
    \hat{\bbx} = \argmin_{\bbx \in \{\bar{\bbx}_m\}_{m=1}^M} ||\bby - \bbH\bbx||_2^2.
\end{equation}

\noindent Notice that these $M$ trajectories can be run in parallel, as they are independent of each other.

\vspace{1mm}
\noindent {\bf Computational complexity.}
The first step in Alg.~\ref{alg} is to compute the SVD of the channel $\bbH$, whose complexity is $\ccalO(N_uN_r\min\{N_u, N_r\})$, and is done only once per channel.
Moreover, notice that the matrices involved in each iteration are diagonal so the complexity of multiplying them is $\ccalO(N_u^2)$, while the matrix inversion is $\ccalO(N_u)$.
Finally, given a modulation of $K$ symbols, the complexity of~\eqref{E:mixed_gaussian} is $\ccalO(KN_u)$.
Hence, one iteration has a complexity of $\ccalO(N_u^2 + KN_u)$.
The overall complexity, including the SVD computation and all the iterations, is $\ccalO(N_uN_r\min\{N_u, N_r\} + LT(N_u^2 + KN_u))$.
Regarding the $M$ trajectories, observe that these are independent of each other, so they can be computed in parallel.
To given some context about the performance, we show in Table~\ref{table:complexity} the complexity of the baseline detectors that are used in the numerical experiments\footnote{For RE-MIMO, each layer corresponds a block of an encoder module given by a Transformer with a fixed number of layers, and a predictor module parameterized by an MLP. See~\cite{remimo} for more details.}. 
We observe that our method has also a polynomial complexity, and has a similar complexity as long as the number of noise levels $L$ and iterations per level $T$ are controlled and pushed to small values.
Therefore, the bottlenecks are twofold: the SVD computation and the number of iterations $LT$. 
While the former is inevitable, the latter is a parameter that we control and represents a trade-off between SER performance and computational complexity.
In Section~\ref{sec:results}, we present numerical experiments that analyze this trade-off and the impact on the SER performance.

\begin{algorithm}[t]
	\caption{Annealed Langevin for MIMO detection}\label{alg}
	\begin{algorithmic}
		\Require $T, \{\sigma_l\}_{l=1}^L, \epsilon, \sigma_0, \bbH, \bby, \tau$
		\State Compute SVD of $\bbH = \bbU\boldsymbol{\Sigma}\bbV^{\top}$
		\State Initialize $\tilde{\boldsymbol{\chi}}_{t=0,l=1}$ with random noise $\ccalU[-1,1]$
		\For{$l = 1\; \text{to}\;  L$}
		\State $[\boldsymbol{\Lambda}_l]_{jj} =$
		$\begin{cases}
		  \frac{\epsilon \sigma_l^2 }{\sigma_L^2} (1 - \frac{\sigma_l^2}{\sigma_0^2}s_j^2) \hspace{8mm} \text{if} \,\,\, \sigma_ls_j \leq \sigma_0 \\
		\frac{\epsilon}{\sigma_L^2} (\sigma_l^2 - \frac{\sigma_0^2}{s_j^2}) \hspace{10mm}  \text{if} \,\,\, \sigma_ls_j > \sigma_0
		\end{cases}$
		
		\For{$t = 0\; \text{to}\; T-1$}
			\State Draw $\bbw_t \sim \ccalN(0, \bbI)$
			
            \State Compute $\nabla_{\tilde{\boldsymbol{\chi}}_{t,l}}\log p(\boldsymbol{\eta}|\tilde{\boldsymbol{\chi}}_{t,l}, \bbH)$ as in~\eqref{eq:score_likeli}
            
            \State Compute $\nabla_{\tilde{\bbx}_{t,l}}\log p(\tilde{\bbx}_{t,l})$
            as in~\eqref{eq:prior}
            
            \State Compute $\nabla_{\tilde{\boldsymbol{\chi}}_{t,l}}\!\log p(\tilde{\boldsymbol{\chi}}_{t,l}| \boldsymbol{\eta}, \bbH)$ as in~\eqref{eq:full_score}

            
            \State $\tilde{\boldsymbol{\chi}}_{t+1, l} \!=\! \tilde{\boldsymbol{\chi}}_{t,l} + \boldsymbol{\Lambda}_l \nabla_{\tilde{\boldsymbol{\chi}}_{t,l}}\!\log p(\tilde{\boldsymbol{\chi}}_{t,l}| \boldsymbol{\eta}, \bbH) + \sqrt{2\boldsymbol{\Lambda}_l \tau}\, \bbw_t$
		\EndFor
		\State $\tilde{\boldsymbol{\chi}}_{0, l+1} = \tilde{\boldsymbol{\chi}}_{T, l}$
		\EndFor \\
	\Return $\bar{\bbx} = \argmin_{\bbx \in \ccalX^{N_u}}||\bbx - \bbV\tilde{\boldsymbol{\chi}}_{T,L}||_2^2$
	\end{algorithmic}
\end{algorithm}

\section{Unfolding of the Langevin detector}\label{sec:unrolling}

In this section, we present an unfolded MIMO detector inspired on the Langevin detector presented in Section~\ref{subsec:langevindyn}.
We briefly introduce the algorithm unfolding framework in Section~\ref{subsec:unrolling} while in Section~\ref{subsec:architecture} we explain our proposed architecture.

\subsection{Algorithm unfolding}
\label{subsec:unrolling}

Algorithm unfolding (or unrolling)~\cite{eldar2021, UWMMSE, farsad2021} refers to the general notion of selecting an iterative model-based approach to solve a problem of interest and building a problem-specific neural network with layers inspired by these iterations.
Based on its recent success on many problems, we propose to apply this framework to the annealed Langevin dynamic explained in Section~\ref{sec:langevin} by learning the score of the likelihood.

In other contexts, several previous works have already proposed to parameterize the score function by a \textit{score network}. 
In~\cite{ermon2019, ermon2020}, the authors proposed to parameterize the score function of the distribution and train using denoising score matching~\cite{denoisingscore}.
In~\cite{kawar2021snips}, instead of working with a general distribution, they focus on sampling from a posterior distribution.
Therefore, they approximate the score of the prior by a neural network, but the training process is done similarly as in~\cite{ermon2019}.
In this work, we leverage the algorithm unfolding framework to train our proposed architecture end-to-end.
In particular, given that we have access to a set of pilot signals, we can train in a supervised way by minimizing the error between our estimation and the true symbol.

\begin{table}[t]
    \centering
    \begin{tabular}{ | c | c | }
        \hline
        Method & Complexity \\ 
        \hline
        MMSE & $\ccalO(N_u^3 + N_r N_u^2)$  \\ 
        \hline
        OAMPNet & $\ccalO(N_r^3 + N_u^3 + N_rN_u^2 + N_uN_r^2)$ per layer \\ 
        \hline
        RE-MIMO & $\ccalO(N_r^2N_u + N_r N_u^2)$ per EP block\\
        \hline
        Langevin & $\ccalO(N_uN_r\min\{N_u, N_r\} + LT(N_u^2 + KN_u))$\\ 
        \hline
    \end{tabular}
    \caption{Complexity of Langevin and the other baseline detectors. 
    We observe that our proposed detector has a polynomial complexity much like the other detectors.}
    \label{table:complexity}
\end{table}

\subsection{Detection by sampling from the posterior distribution using a learned score network.}
\label{subsec:architecture}

The main objective of the proposed learning-based detector is to be robust against noise variance uncertainty while following the same structure as the detector in Section~\ref{subsec:langevindyn}. 
In general, learning-based detectors need access to the true noise variance as they use this parameter in the denoising step~\eqref{eq:general_iter_2}.
However, in our proposed detector we avoid this denoising step by following the annealed Langevin dynamic.
Therefore, the only noise variance-dependent term in our detector is the score of the likelihood in~\eqref{eq:score_likeli}, so we propose to parameterize it by a \textit{score network}.
We detail now the architecture of our proposed learning-based detector.

\begin{figure*}[t]
	\centering
	\includegraphics[width=0.9\textwidth]{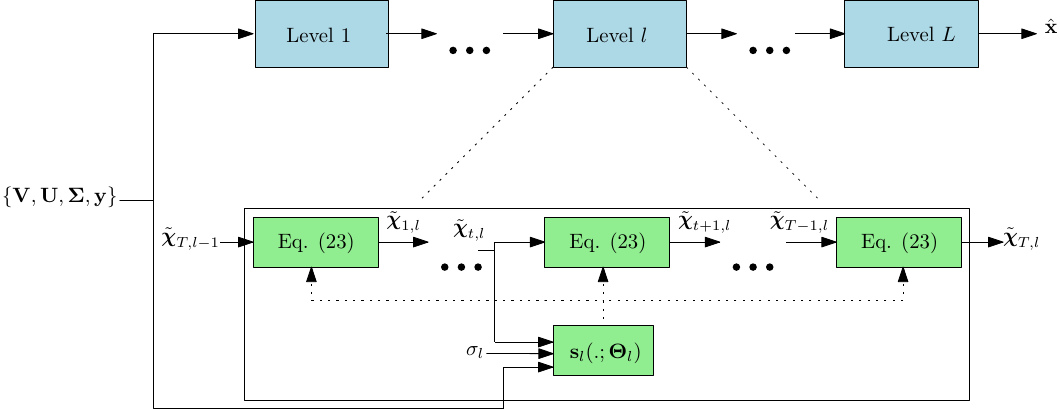}
	\vspace{-0.03in}
	\caption{{\small Scheme of our proposed unfolded Langevin detector. 
	The input to architecture is the SVD of $\bbH$ and the observation $\bby$ and the estimated symbol is computed by running sequentially the Langevin dynamic for each of the $L$ levels, represented by blue.
	Inside each level $l$, we run~\eqref{eq:full_score_unfold} for $T$ iterations,	where the score network $\bbs_l(.; \boldsymbol{\Theta})$ is the same for all the interatios in the level.
	This is represented by green, and the output $\tilde{\boldsymbol{\chi}}_{T, l}$ is the initial value for the next level $l-1$.}}
	\vspace{-0.1in}
	\label{fig:architecture}
\end{figure*}

\vspace{1mm}
\noindent {\bf Architecture.} Our unfolded architecture (depicted in Fig.~\ref{fig:architecture}) is composed of $L$ noise levels as the detector derived in Section~\ref{sec:langevin}, where the score of the likelihood is replaced by a \textit{score network}.
Formally, we learn a function $\Phi: \mathbb{C}^{N_r} \times \mathbb{C}^{N_r \times N_u} \rightarrow \ccalX^{N_u}$ such that $\Phi(\bbH, \bby; \boldsymbol{\Theta})$ is a solution of~\eqref{eq:ml}.
Each layer $l$ of $\Phi(\bbH, \bby; \boldsymbol{\Theta})$, defined as $\phi_l(\bbH, \bby, \tilde{\boldsymbol{\chi}}_{T, l-1}; \boldsymbol{\Theta}_l)$ where $\boldsymbol{\Theta} = \{\boldsymbol{\Theta}_l\}_{l=1}^L$, is composed by T iterations and its output is the initial value for the next layer, i.e., $\tilde{\boldsymbol{\chi}}_{T, l}$.
Thus, each iteration is defined as
\begin{align}\label{eq:full_score_unfold}
    \tilde{\boldsymbol{\chi}}_{t+1, l} \!=\! \tilde{\boldsymbol{\chi}}_{t,l} + \frac{\epsilon \sigma_l^2 }{\sigma_L^2} \!\Big(\bbs_l(\boldsymbol{\eta} - \boldsymbol{\Sigma}\tilde{\boldsymbol{\chi}}_{t,l}, \tilde{\boldsymbol{\chi}}_{t,l}, \sigma_l; \boldsymbol{\Theta}_l) \\
    + \bbV^{\top}\nabla_{\tilde{\bbx}_{t,l}}\log p(\tilde{\bbx}_{t,l})\Big) + \sqrt{2\frac{\epsilon \sigma_l^2 }{\sigma_L^2} \tau}\, \bbw_t.
\nonumber
\end{align}

\noindent The neural network $\bbs_l(.; \boldsymbol{\Theta}_l)$ is the \textit{score network} at level $l$ and $\boldsymbol{\Theta}_l$ is the corresponding set of learning parameters.
Given that the score of the likelihood~\eqref{eq:score_likeli} depends on the observations $\bby$, the channel $\bbH$, the sample $\tilde{\boldsymbol{\chi}}_{t,l}$ at iteration $t$ and the noise variance of the annealed noise $\sigma_l$, we define the tuple $\{\boldsymbol{\eta} - \boldsymbol{\Sigma}\tilde{\boldsymbol{\chi}}_{t,l}, \tilde{\boldsymbol{\chi}}_l, \sigma_l\}$ as the input of the score network.
We decided to work in the spectral domain based on the experimental results, which were better than using $\bbH$ directly.

In essence, our architecture only replaces the noise variance-dependent term, which is the score of the likelihood, by a neural network, and leverage the score of the prior to generate samples from the constellation without a projection step.
Each score network $\bbs_l$ is a multi-layer perceptron (MLP) with 4 layers: the first has 400 hidden neurons, the second one 350 and the last one 100.
With the exception of the last layer, all the others use an exponential linear unit (ELU) function as point-wise non-linearity.

\vspace{1mm}
\noindent {\bf Loss function.} Our method concatenates $L$ layers, where each layer has its own learning parameters. 
We use the average $\ell_2$ loss over all the layers. 
More precisely, the loss function is defined as 
    \begin{equation}\label{eq:loss}
        \ccalL(\boldsymbol{\Theta}) = \mathbb{E}_{\bbx, \bby, \bbH}\bigg[\frac{1}{L}\sum_{l=1}^L||\phi_l(\bbH, \bby, \tilde{\boldsymbol{\chi}}_{T, l-1}; \boldsymbol{\Theta}_l) - \bbx||_{\ell_2}^2\bigg],
    \end{equation}
and we train the neural networks $\bbs_l(.; \boldsymbol{\Theta}_l)$ (implicit in \eqref{eq:loss} within $\phi_l$) by minimizing the empirical loss. 

\vspace{1mm}
\remark{In the numerical results we consider the learning architecture with Gaussian noise. 
However, no assumptions are made with respect to the noise distribution.
Hence, the method can be used also in non-Gaussian cases; the only requirement is having an estimation of the likelihood model, which can be obtained as in~\cite{normalizingflows}.}

\section{Results}\label{sec:results}
In this section we present the results of our proposed method.\footnote{Code to replicate the numerical experiments can be found
at \url{https://github.com/nzilberstein/Langevin-MIMO-detector}.}
We start by analyzing the SER performance of our detector when considering different noise levels $L$, different number of trajectories $M$ and different temperature parameter $\tau$.
Based on this analysis, we fix the optimal hyper-parameters and we compare our method with both classical and learning-based detectors for different channel models -- both simulated and real ones -- and settings.
We also evaluate our method when users are transmitting with different modulation schemes simultaneously. 
Finally, the robustness of our unfolded method presented in Section~\ref{sec:unrolling} is compared with other detectors when there is uncertainty in the noise.

Before moving to the experiments, we first discuss the implementation details of the baseline detectors that are used in the experiments.
For the learning-based ones, we briefly explain the training process.

\begin{itemize}
    \item {\bf MMSE}~\cite{Proakis2007}: Linear detector, which formally solves first a relaxation of~\eqref{P:main}, by defining $\bbx \in \ccalX^{N_u}$ as $\bbx \in \mathbb{C}^{N_u}$, and projection onto one of the constellation elements.
    \item {\bf V-BLAST}~\cite{Chin2002ParallelMD}: Multi-stage interference cancellation BLAST algorithm using Zero-Forcing as the detection stage.
    \item {\bf MMNet-iid}~\cite{mmnet}: Learning based detector that is channel-dependent. 
    Hence, we use as a baseline only for the Rayleigh channel model, as in this case it becomes channel-agnostic.
    We implement the scheme that has only two scalar parameters per layer with 10 layers~\cite{mmnet}.
    \item {\bf RE-MIMO}~\cite{remimo}: Recurrent permutation equivariant neural detector based on an encoder-predictor architecture. The encoder is parameterized by a transformer, while the predector with an MLP.
    We train as proposed in the paper.
    \item {\bf OAMPNet}~\cite{oampnet}: We use 10 layers as it is proposed in the paper. 
    At each layer, a matrix pseudo inverse is required and has 2 learnable parameters.
    \item {\bf Langevin detector}: Detector proposed in Section~\ref{sec:langevin}.
    The following hyper-parameters are fixed for all the experiments unless otherwise mentioned: $\epsilon$ is fixed at $3 \times 10^{-5}$, the number of samples per noise level at $T=70$, the number of noise levels at $L = 20$ between $\sigma_1 = 1$ and $\sigma_{20} = 0.01$. 
    The temperature parameter is fixed at $\tau = 1/2$ and the number of trajectories to $M = 20$.
    \item {\bf U-Langevin detector}: Detector based on an unfolding of the annealed Langevin dynamic as defined in Section~\ref{sec:unrolling}.
    The training is performed using Adam optimizer \cite{kingma2017adam} for $15,\!000$ iterations with a batch size of $500$ matrices per iteration. 
    The learning rate that starts at $l_r = 10^{-3}$ and is scaled by a factor of $10^{-1}$ after each 5000 iterations.
    We consider the same hyperparameter settings as {\bf Langevin detector}, but with $\tau = 1$.
    \item {\bf ML}: The optimal solver for~\eqref{eq:ml} using a highly-optimized mixed integer programming package Gurobi~\cite{gurobi}.
\end{itemize}

In all experiments, the signal-to-noise ratio (SNR) is given by 
\begin{equation}
\label{eq:SNRtrue}
	\text{SNR} = \frac{\mathbb{E}[||\bbH\bbx||^2]}{\mathbb{E}[||\bbz||^2]}.
\end{equation}

\noindent For all the experiments in the following subsections, we consider two different modulation schemes: 16-QAM and 64-QAM.
The simulation environment includes a base station with $N_r=64$ receiver antennas and $N_u=32$ single-antenna users, except for the real channel, which is further detailed in Section~\ref{subsec:realchannel}.

\subsection{Analysis of Langevin method}

\begin{figure*}[t]
    \centering
	\begin{subfigure}{.4\textwidth}
    	\centering
    	\includegraphics[width=1\textwidth]{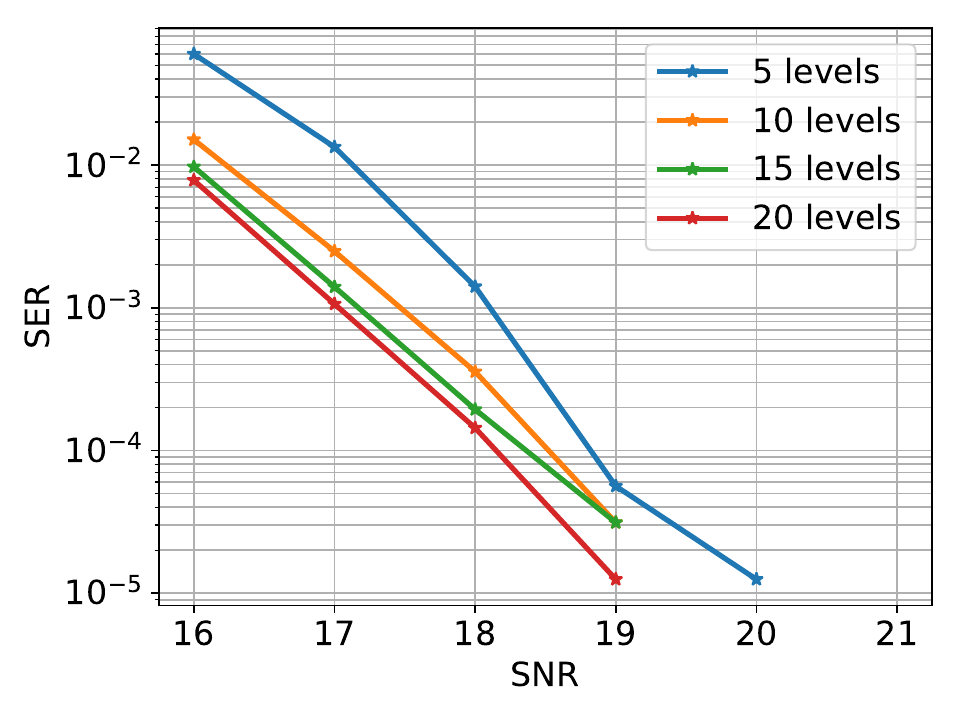}
    	\vspace{-0.15in}
    	\caption{}
    	\label{fig:SER-noiselevels}
	\end{subfigure}%
	\centering
	\begin{subfigure}{.4\textwidth}
    	\centering
    	\includegraphics[width=1\textwidth]{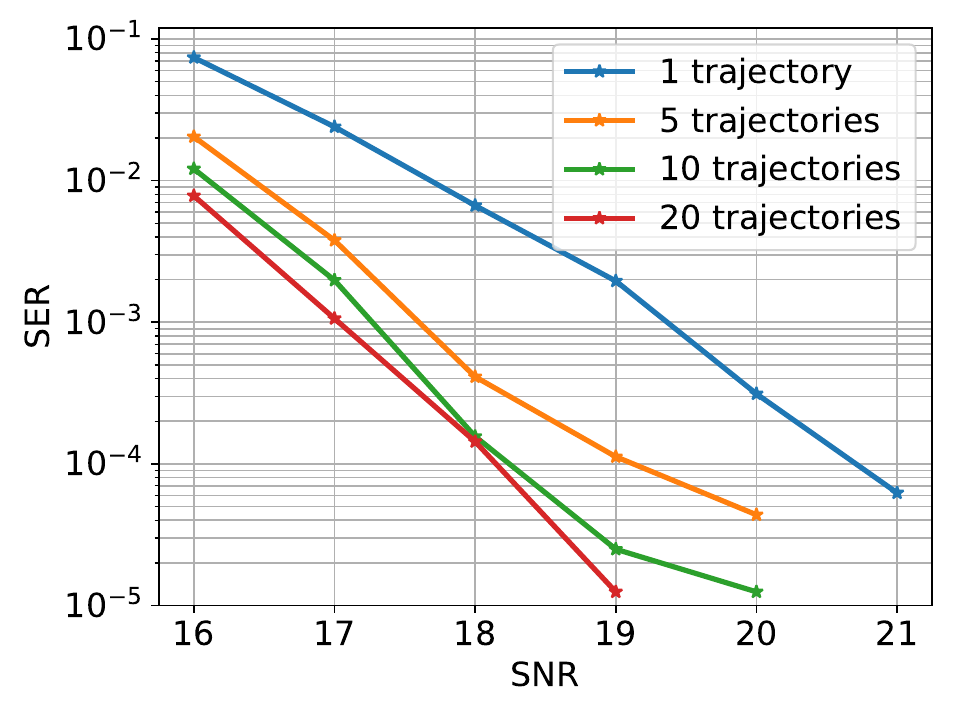}
    	\vspace{-0.15in}
    	\caption{}
    	\label{fig:SER-sertraj}
	\end{subfigure}
	\centering
	\begin{subfigure}{.4\textwidth}
    	\centering
    	\includegraphics[width=1\textwidth]{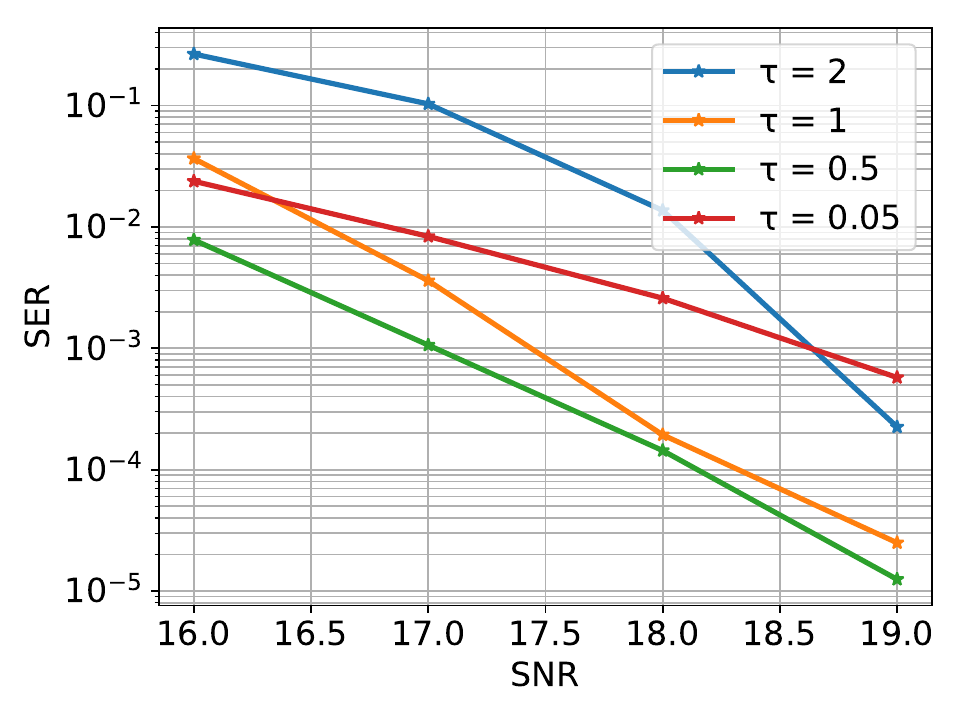}
    	\vspace{-0.15in}
    	\caption{}
    	\label{fig:SER-temperature}
	\end{subfigure}
    \centering
    \begin{subfigure}{.4\textwidth}
        \centering
        \includegraphics[width=1\textwidth]{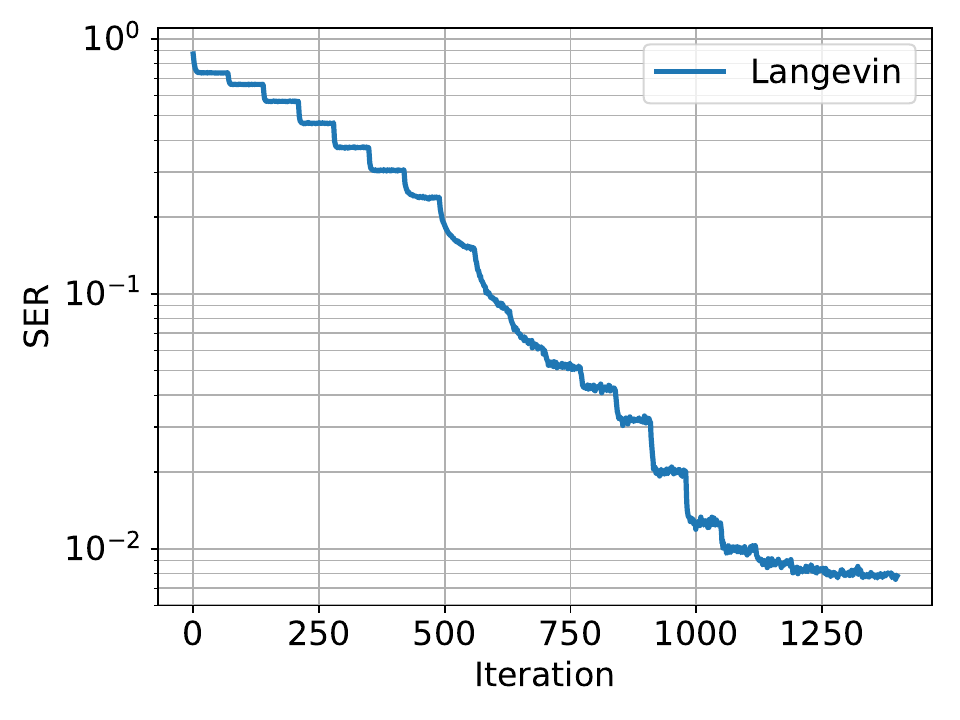}
        \vspace{-0.15in}
        \caption{}
        \label{fig:iter}
    \end{subfigure}
	\vspace{-0.02in}
	\caption{ {\small Performance analysis of our proposed method evaluated in a Kronecker correlated channel model as in~\eqref{E:kron}. (a)~SER as a function of SNR for our Langevin method for noise levels $L \in \{5, 10, 15, 20\}$. (b)~SER as a function of SNR for our Langevin method for $M \in \{1, 5, 10, 20\}$ numbers of trajectories. (c)~SER as a function of SNR for two different temperature parameters $\tau$. (d)~Numerical convergence of our proposed detector, in terms of SER as a function of iterations.} }
	\vspace{-0.1in}
	\label{figs_analysis_detector}
\end{figure*}

In this subsection we consider the Kronecker correlated channel model, which is generated following
\begin{equation}
	\bbH = \bbR_r^{1/2}\bbH_e \bbR_u^{1/2},
	\label{E:kron}
\end{equation}
where $\bbH_e$ is a Rayleigh fading channel matrix and $\bbR_r$ and $\bbR_u$ are the spatial correlation matrices at the receiver and transmitters, respectively.
These correlation matrices are generated according to the exponential correlation matrix model with a coefficient $\rho = 0.6$; see \cite{Loyka2001} for details. 

\vspace{3mm}
\noindent{\bf Varying the number of noise levels.} 
In the first experiment, given the sequence of noise levels with variance $\{\sigma_l\}_{l=1}^L$, we fix $\sigma_1 = 1$, $\sigma_L = 0.01$, $M=20$ trajectories and a temperature $\tau=1/2$ and change the number of noise levels between them.
We consider four cases where $L \in \{5, 10, 15, 20\}$; see Fig.~\ref{fig:SER-noiselevels}.
First, notice that the performance, for low SNR when $L=5$, is much worse than the other three cases.
This implies that the algorithm is not able to sufficiently explore the search space. 
A natural trade-off between computational burden and performance exists: when considering more levels, the SER performance improves at the cost of increasing the running time.
From this experiment, we conclude that at least $L=10$ levels are needed in order to perform as well as the existing state-of-the-art detectors.

\vspace{3mm}
\noindent{\bf Varying the number of trajectories.} 
The performance of the detector as a function of the number $M$ of different Langevin trajectories [cf.~\eqref{eq:Ntraj}] is shown in Fig.~\ref{fig:SER-sertraj}.
We analyze five cases where $M \in \{ 1, 5, 10, 20\}$.
For all the cases, we consider $L=20$ noise levels and $\tau=1/2$.
From the results, we see that $M$ is a hyperparameter that has a high impact on the overall performance of the detector. 
In particular, if we consider only $M=1$, then the performance degrades severally, with a SER, for low SNR, in the order of the classical MMSE detector (not shown in the figure).
However, if we consider $M=20$ trajectories, the proposed method outperforms state-of-the-art detectors, as we illustrate in our next experiment.

\vspace{3mm}
\noindent{\bf Varying the temperature parameter.} 
In this experiment, given a fixed sequence of noise levels and $M=20$ trajectories, we change the temperature parameter $\tau$.
We consider four cases where $\tau \in \{2, 1, 0.5, 0.05\}$; see Fig.~\ref{fig:SER-temperature}.
When $\tau = 1$, we are sampling from the original target distribution as it was described in Section~\ref{sec:langevin}.
While increasing the temperature parameter to $2$ increase the mixing time, when the value of $\tau$ is decreased to $0.5$ the mixing time improves, although we are not sampling from the original posterior distribution. 
This is related to the landscape of the posterior distribution, as the temperature is scaling the local maxima and concentrating around the global maxima.
Lastly, notice that decreasing further the value of $\tau$ has a negative impact, as the mixing time increases.

\vspace{3mm}
\noindent{\bf Convergence result.} 
The last experiment of this subsection is a numerical result for the convergence of our method, where we compute the SER as a function of the iterations.
We fixed the parameters to the best combination: a fixed sequence of $20$ noise levels, $M=20$ trajectories, and a temperature parameter $\tau = 0.5$.
Moreover, we consider an SNR = \SI{16}{\decibel}.
The result is shown in Fig.~\ref{fig:iter}.
The total number of iterations is $\text{number of noise levels} \times \text{number of samples per level}$, which for our case is $20 \times 70 =1400$.
Therefore, as there 20 levels of noise, there are also 20 jumps between each level in SER.
Moreover, the figure reveals that the first and the last levels could be shorter without affecting the performance, as the SER at those levels is almost constant. 
Thus, the complexity of the detector can be reduced by using an adaptive number of iterations at each level.
Although in this work we consider a fixed number of iterations per level for simplicity, we leave the exploration of this direction as future work.

\subsection{Comparison with other methods}
\label{Ss:baseline}

In this subsection we compare our detector with baseline methods.
We consider three different channels: Rayleigh MIMO channel, Kronecker correlated MIMO channel and 3GPP channel.

\begin{figure*}[t]
    \centering 
	\begin{subfigure}{.4\textwidth}
    	\centering
    	\includegraphics[width=1\textwidth]{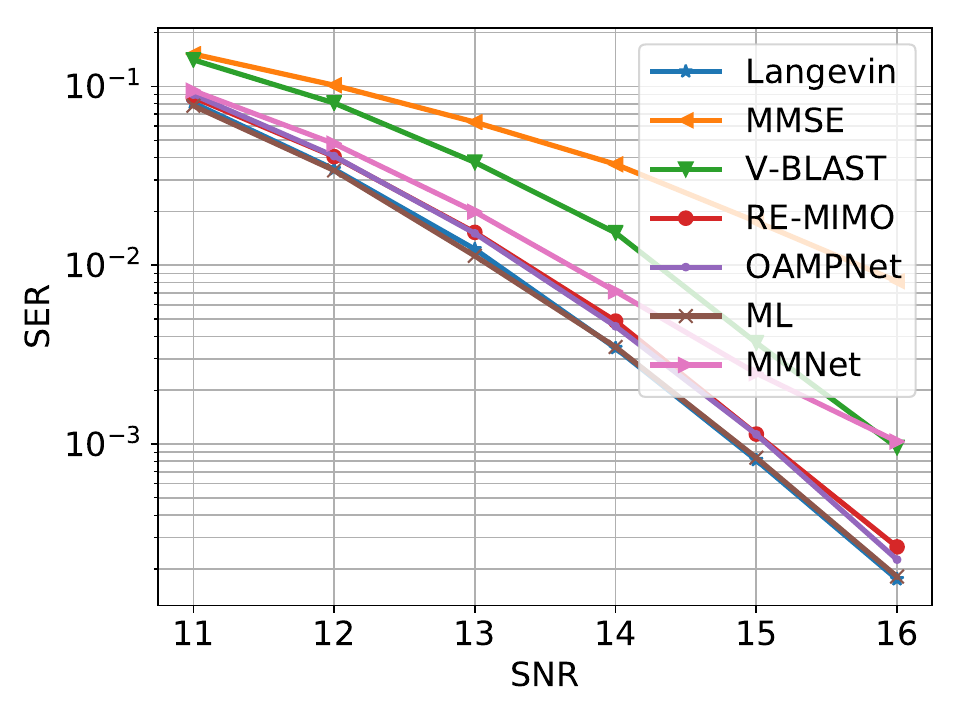}
    	\vspace{-0.15in}
    	\caption{}
    	\label{fig:SER-iidchannel}
	\end{subfigure}%
	\begin{subfigure}{.4\textwidth}
    	\centering
    	\includegraphics[width=1\textwidth]{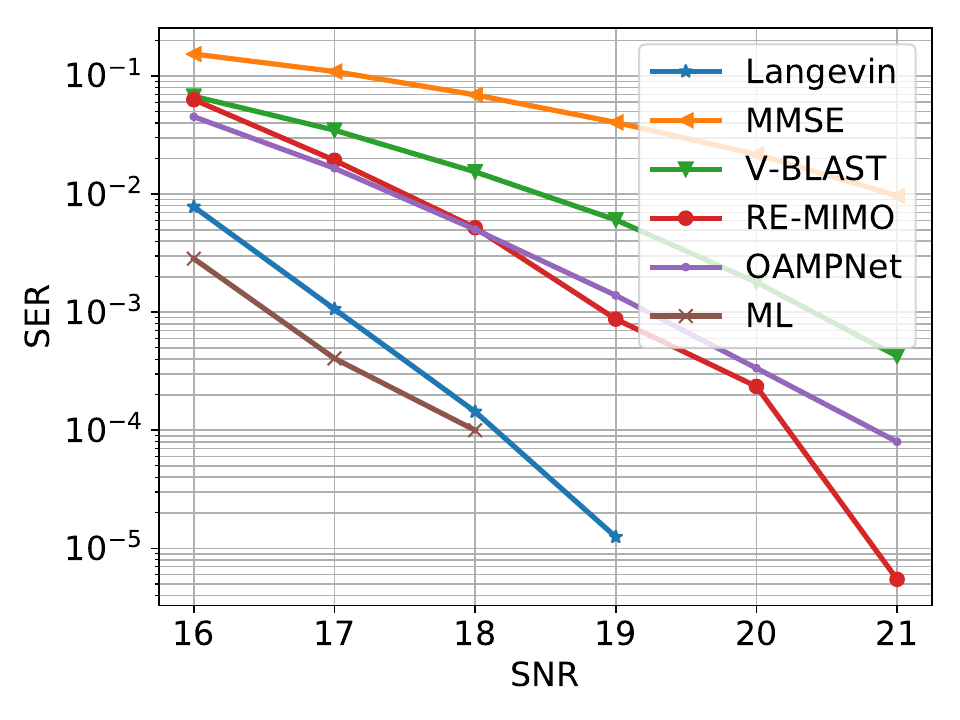}
    	\vspace{-0.15in}
    	\caption{}
    	\label{fig:SER-comparison16qamkron}
	\end{subfigure}
	\begin{subfigure}{.4\textwidth}
    	\centering
    	\includegraphics[width=1\textwidth]{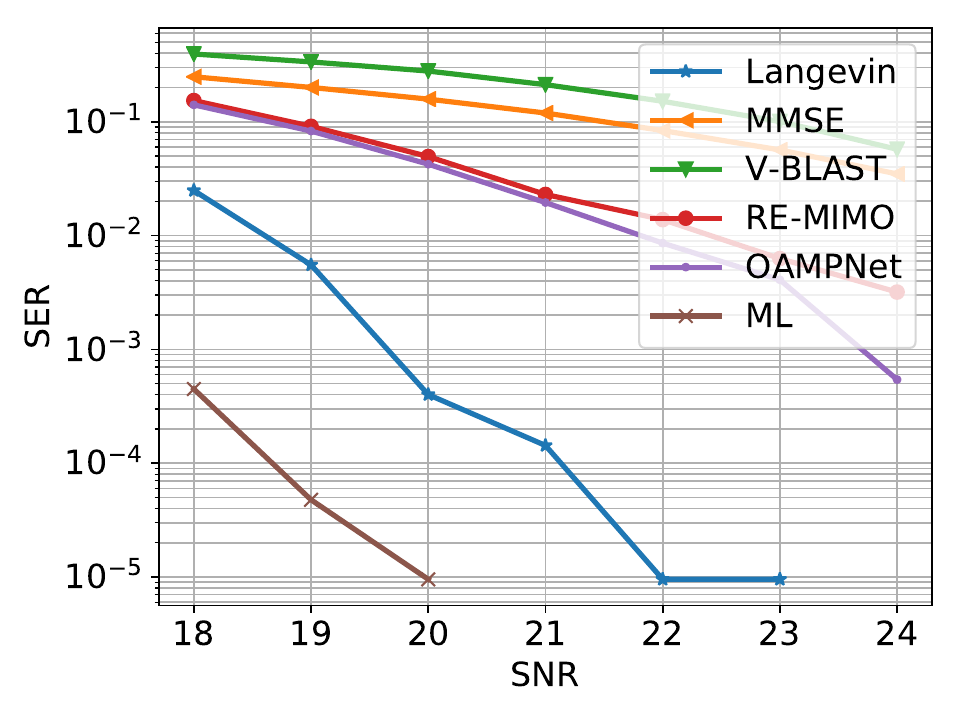}
    	\vspace{-0.15in}
    	\caption{}
    	\label{fig:SER-comparison3gpp16}
	\end{subfigure}
	\begin{subfigure}{.4\textwidth}
    	\centering
    	\includegraphics[width=1\textwidth]{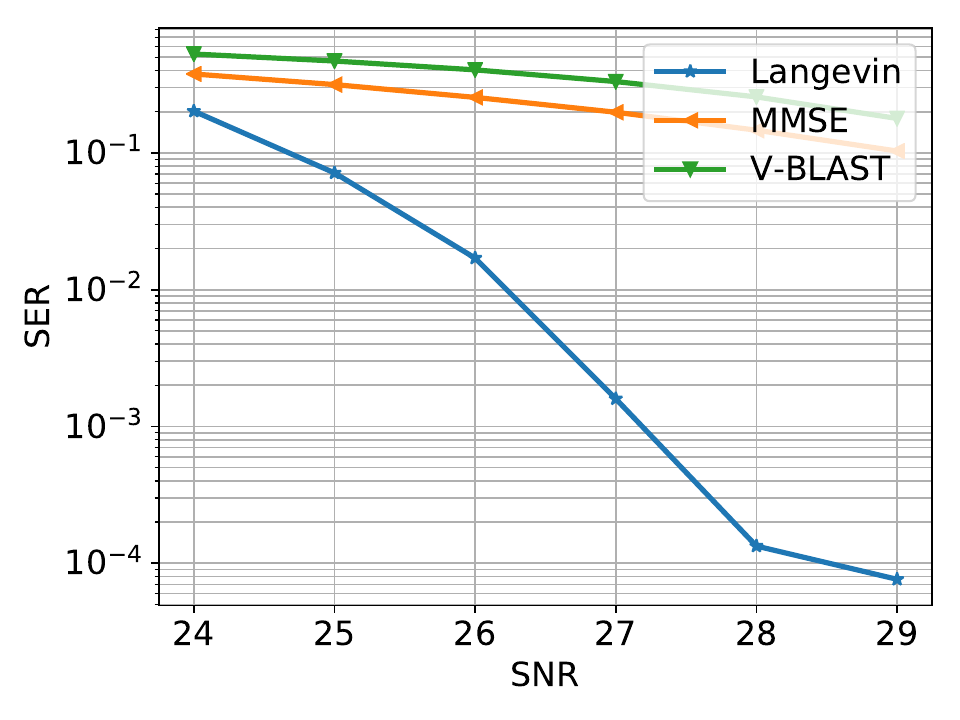}
    	\vspace{-0.15in}
    	\caption{}
    	\label{fig:SER-comparison3gpp64}
	\end{subfigure}%
	\vspace{-0.02in}
	\caption{ {\small SER as a function of SNR for different detection methods evaluated in (a)~Rayleigh channel model and a modulation 16-QAM. (b)~Kronecker correlated channel model as in~\eqref{E:kron} and a modulation 16-QAM. (c)~3GPP channel model and a modulation 16-QAM. (d)~3GPP channel model and a modulation 64-QAM.}}
	\vspace{-0.1in}
	\label{fig_results_synthethic_chann}
\end{figure*}

\vspace{3mm}
\noindent {\bf Rayleigh MIMO channel model.} In this first case, we consider i.i.d. Gaussian channels. 
Hence, each element of $\bar{\bbH}$ is sampled from $\bar{h}_{ij} \sim \mathcal{CN}(0, 1/N_r)$, and then converted to the equivalent real representation. 


The comparison for a batch size of $5000$ channel realizations is shown in Fig.~\ref{fig:SER-iidchannel}. 
The figure shows that our method not only outperforms the other detectors, both the learning-based and the classical ones, but also achieves the optimal ML performance.
In particular, for SNR = \SI{11}{\decibel} Langevin achieves an SER of $8.1\times10^{-2}$, slightly better than its best competitor, RE-MIMO, which achieves an SER of $8.6\times10^{-2}$, while for SNR = \SI{16}{\decibel} Langevin achieves an SER of $1.7\times10^{-4}$ while the best competitor, OAMPNet, an SER of $2.2\times10^{-4}$.
Furthermore, the gap between Langevin and MMSE increases as the SNR increases.
It is particularly interesting to notice that our proposed method outperforms the learning-based detectors, which have been trained with channel instances drawn from the same model as those in the testing set.

\vspace{3mm}
\noindent {\bf Correlated MIMO channel model.} In this second case, we consider the channel defined in~\eqref{E:kron}. Again, the batch size for testing is $5000$ and $\rho = 0.6$. 

The comparison is shown in Fig.~\ref{fig:SER-comparison16qamkron} for 16-QAM.
The figure reveals that our proposed method markedly outperforms the other detectors.
At is explained in~\cite{mmnet}, in the case of correlated channels, MMNet needs to be trained online for each channel realization correlated channels. 
This, we do not consider.
Again, as in the case of Rayleigh channel, our proposed method outperforms the learning-based detectors, which have been trained with channel instances drawn from the same model as those in the testing set.

\vspace{3mm}
\noindent {\bf 3GPP channel model.}  The last channel model of this subsection is representative of 3GPP 3D MIMO channel model~\cite{3gpp}, 
as implemented in the QuaDRiGa channel simulator~\cite{quadriga}.
We consider a base station with an $8\times8$ half-wavelength space, single-polarization antenna array at a height of \SI{20}{\meter}.
We assume that the BS cover a sector of radius \SI{500}{\meter} and 32 single-polarization omni-directional antennas users are dropped randomly.
Moreover, users are NLOS and indoors.
The carrier frequency is \SI{3.5}{\giga\hertz}, and then each subcarrier has a \SI{100}{\mega\hertz} bandwidth. The spacing within subcarriers is \SI{30}{\kilo\hertz}.
We evaluate the performance in a batch size of 3276 matrices.

The comparison is shown in Figs.~\ref{fig:SER-comparison3gpp16} and~\ref{fig:SER-comparison3gpp64} for a 16-QAM and 64-QAM respectively.  
For the 64-QAM we only consider the non-learning based method that are feasible for a real application, as we proved that Langevin outperforms the learning-based.
We see that our proposed method still outperforms both classical and learning-based detectors by several orders of magnitude.
However, the gap between the ML detector and Langevin detector increases further than for the previous channel models.
Notice that we have to increase the SNR ranges by \SI{7}{\decibel} and \SI{2}{\decibel} with respect to the i.i.d. channel and Kronecker correlated channel respectively, to achieve a similar SER performance.

\subsection{Real-world data and measured channel}
\label{subsec:realchannel}

The real channel is an indoor 64-antenna RENEW massive MIMO channel~\cite{RENEW1}. 
We set up 8 users in the vicinity of the receivers. 
We use the RENEWLab real-time software\footnote{The implementation can be found at \url{https://github.com/renew-wireless/RENEWLab}} to collect datasets in all user configurations and the 802.11 LTS signal as the pilot symbol which is used for channel estimation and beamforming matrix calculation. 
The format of the data OFDM symbols also follows the 802.11 standard where the FFT size is 64. Out of the 64 subcarriers, 48 subcarriers carry data, 4 are pilot subcarriers, and the remaining are null subcarriers. The pilot subcarriers are used for residual phase offset correction. 
We use QPSK as the modulation scheme for our experiments.

\begin{figure*}[t]
    \centering
	\begin{subfigure}{.4\textwidth}
    	\centering
    	\includegraphics[width=1\textwidth]{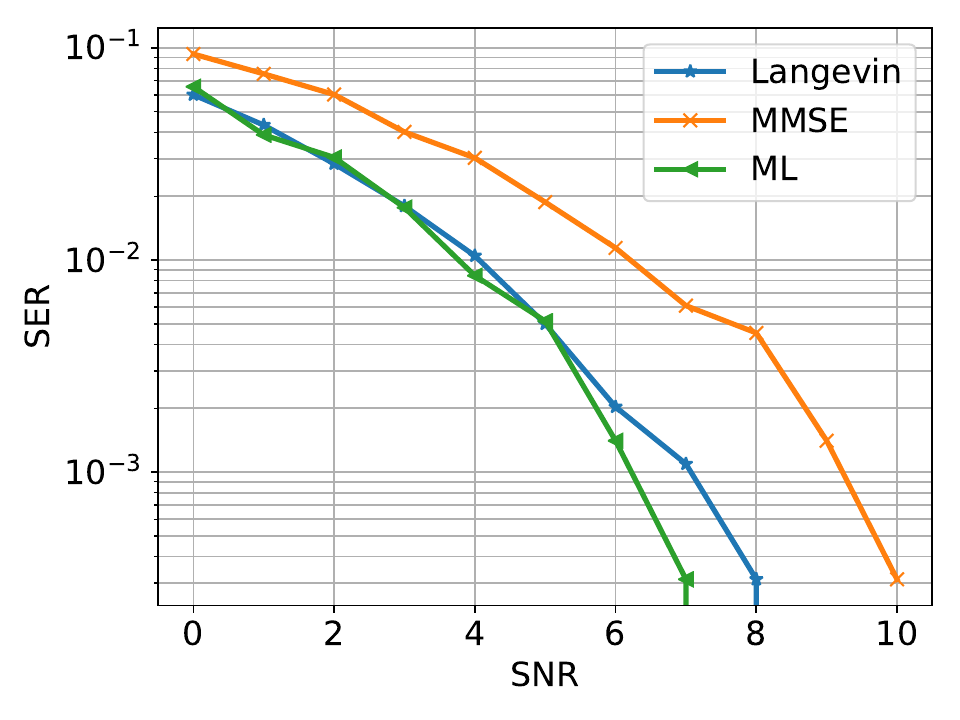}
    	\vspace{-0.15in}
    	\caption{}
    	\label{fig:SER-realH}
	\end{subfigure}
	\begin{subfigure}{.4\textwidth}
    	\centering
    	\includegraphics[width=1\textwidth]{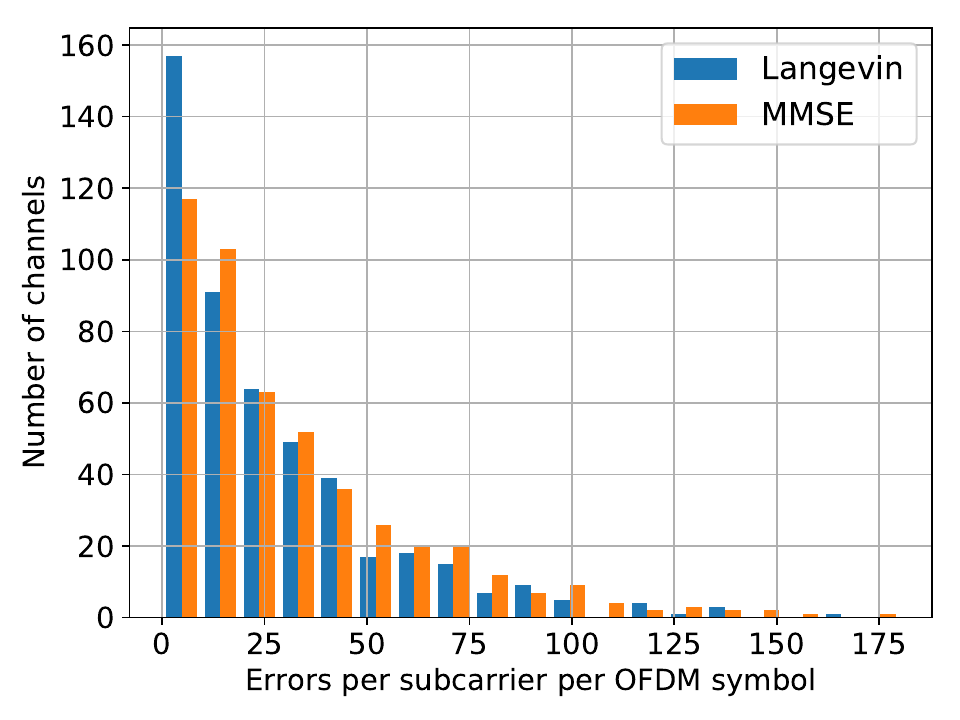}
    	\vspace{-0.15in}
    	\caption{}
    	\label{fig:histogram}
	\end{subfigure}
	\begin{subfigure}{.4\textwidth}
    	\centering
    	\includegraphics[width=1\textwidth]{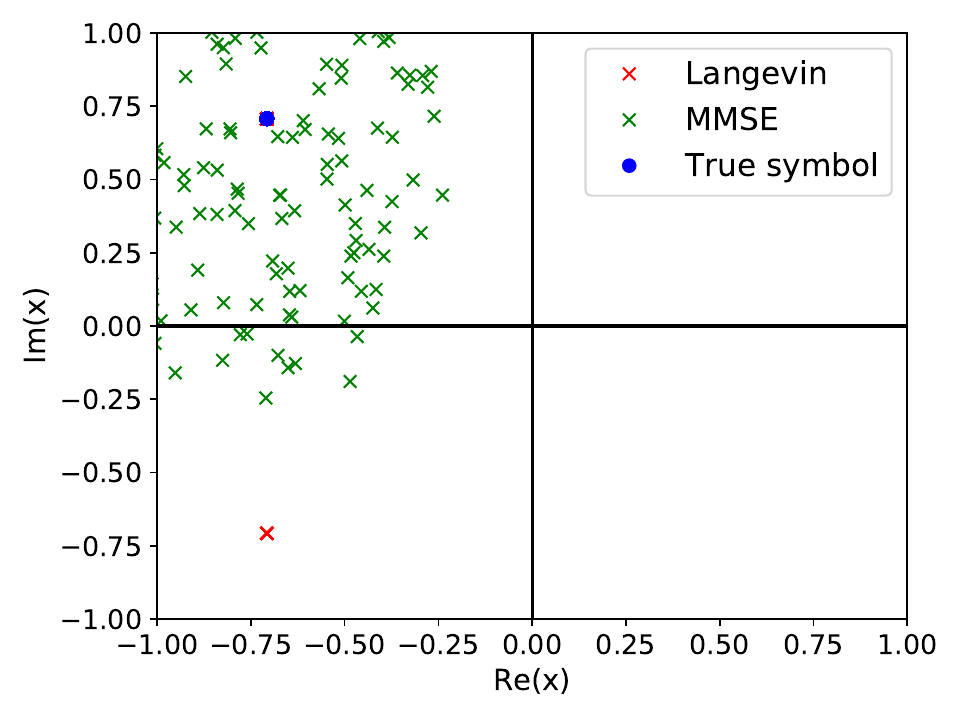}
    	\vspace{-0.15in}
    	\caption{}
    	\label{fig:user0}
	\end{subfigure}
	\begin{subfigure}{.4\textwidth}
    	\centering
    	\includegraphics[width=1\textwidth]{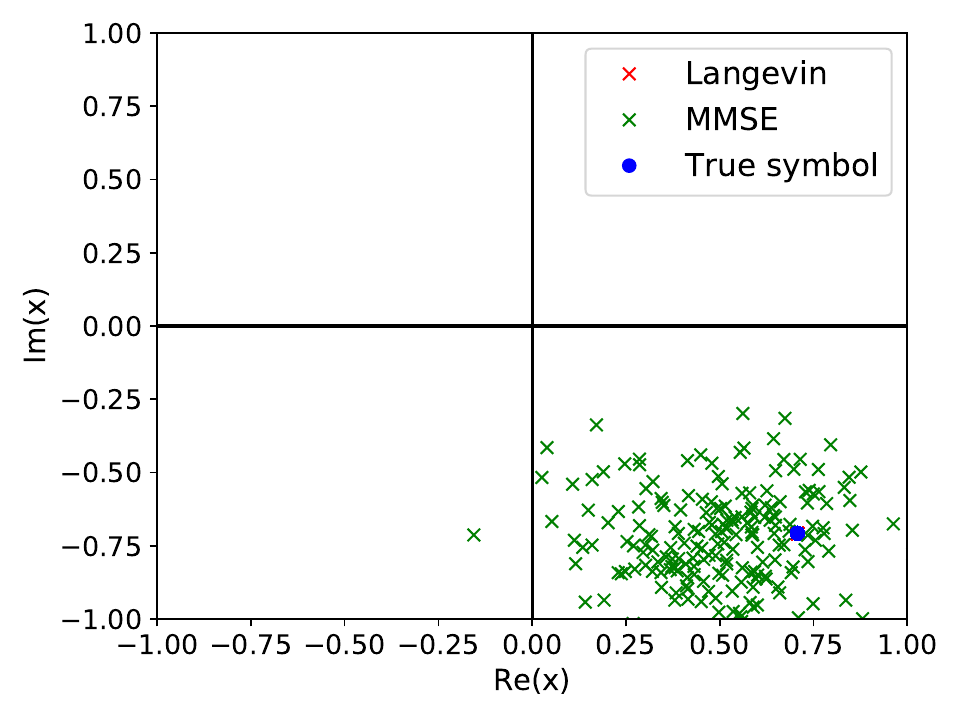}
    	\vspace{-0.15in}
    	\caption{}
    	\label{fig:user7}
	\end{subfigure}
	\vspace{-0.02in}
	\caption{ {\small (a)~~SER as a function of SNR for our Langevin method using the real channel defined in Section~\ref{subsec:realchannel} and with synthetic data. The modulation is $4$-QAM. (b)~Histogram of the number of errors per subcarrier and OFDM symbol for each of the 200 time and frame and 8 users. (c)-(d)~Symbol's estimation for two users and 200 different observations. We consider one OFDM symbol and one subcarrier for both users, and send the same symbol in the 200 time frames.}}
	\vspace{-0.1in}
	\label{fig_results_realchannel}
\end{figure*}

The first experiment is shown in Fig.~\ref{fig:SER-realH} for a QPSK constellation. 
In this case, we generate synthetic observations using a batch of 800 channels.
We compare with MMSE and ML.
We see that our proposed Langevin detector outperforms MMSE and achieves a performance similar to ML.

The second experiment was carried out with real observations, with an SNR$\approx$\SI{20}{\decibel} (this is an average value across all the dataset). 
For this experiment, we change the hyper-parameters of our proposed detector based on an empirical analysis.
In particular, we set $M=2$ trajectories and we increase the number of noise levels to 25 and set $\sigma_L = 0.001$.
In Figs.~\ref{fig:user0} and~\ref{fig:user7} we show for $200$ observations (time frames) for two users, considering one OFDM symbol and one subcarrier.
We consider the 8 users and calculate the number of errors for each OFDM symbol and one subcarrier. 
We compute the histogram of this result, which is shown in Fig.~\ref{fig:histogram}, where each point corresponds to the number of error in $200 \times 8$ observations per OFDM symbol and subcarrier.
From a total number of $960$k observations, 12880 symbols are incorrectly detected by Langevin, while 15212 by MMSE, achieving a SER of 0.016 and 0.02 respectively.
Overall, compared to the synthetic data, the performance of both methods is worse. 
Moreover, although the gap in performance reduces between both methods, notice that for high SNR both perform similarly.

\subsection{Performance for multiple modulation schemes simultaneously}

In this subsection we consider the case of users transmitting with different modulation schemes simultaneously. 
In particular, we assume two different modulations.
Formally, given two constellation sets $\ccalX_1$ and $\ccalX_2$, we define the following forward model

\begin{equation}
    \bby = \bbH \begin{bmatrix}\bbx_1 \\ \bbx_2\end{bmatrix} + \bbz,
\end{equation}
where $\bby \in \mathbb{C}^{N_r}$, $\bbH \in \mathbb{C}^{N_r \times N_t}$ and $\bbx_1\in \ccalX_1^{N_u/2}, \bbx_2 \in \ccalX_2^{N_u/2}$ are two vectors of transmitted symbols corresponding to one of the two modulations.
For the detection, we assume that we know the modulation scheme used by each user.
We use the Kronecker correlated channel model~\ref{E:kron} and the same simulation setting as in Section~\ref{Ss:baseline}.

\begin{figure*}[t]
    \centering
    
    \begin{subfigure}{.4\textwidth}
    	\hspace{-1cm}{\includegraphics[width=1.1\textwidth]{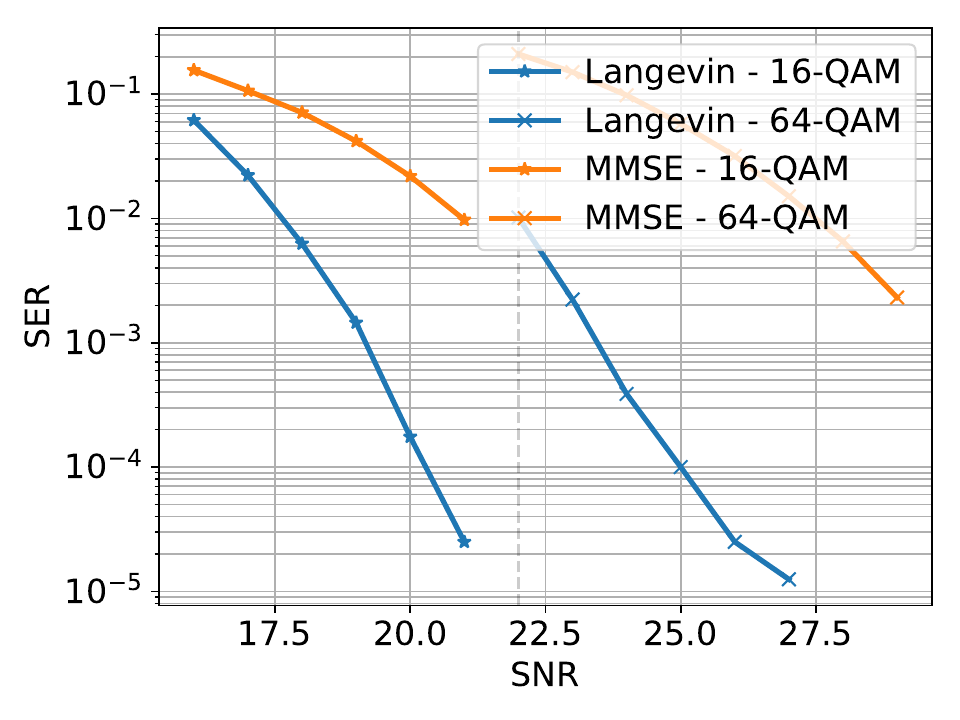}}
    	\vspace{-0.15in}
    	\caption{}
    	\label{fig:SER-twomod}
	\end{subfigure}
	\centering
	\begin{subfigure}{.4\textwidth}
    	\hspace{-0.5cm}{\includegraphics[width=1\textwidth]{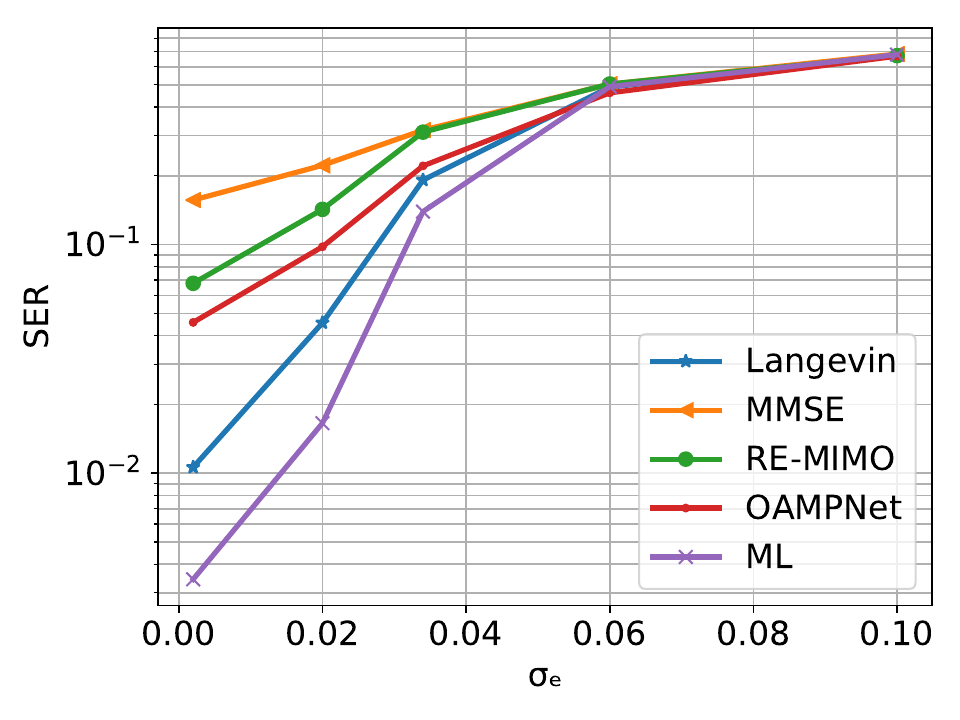}}
    	\vspace{-0.15in}
    	\caption{}
    	\label{fig:imperfectCSI}
	\end{subfigure}
	\centering
	\begin{subfigure}{.4\textwidth}
    	\hspace{-0.5cm}{\includegraphics[width=1\textwidth]{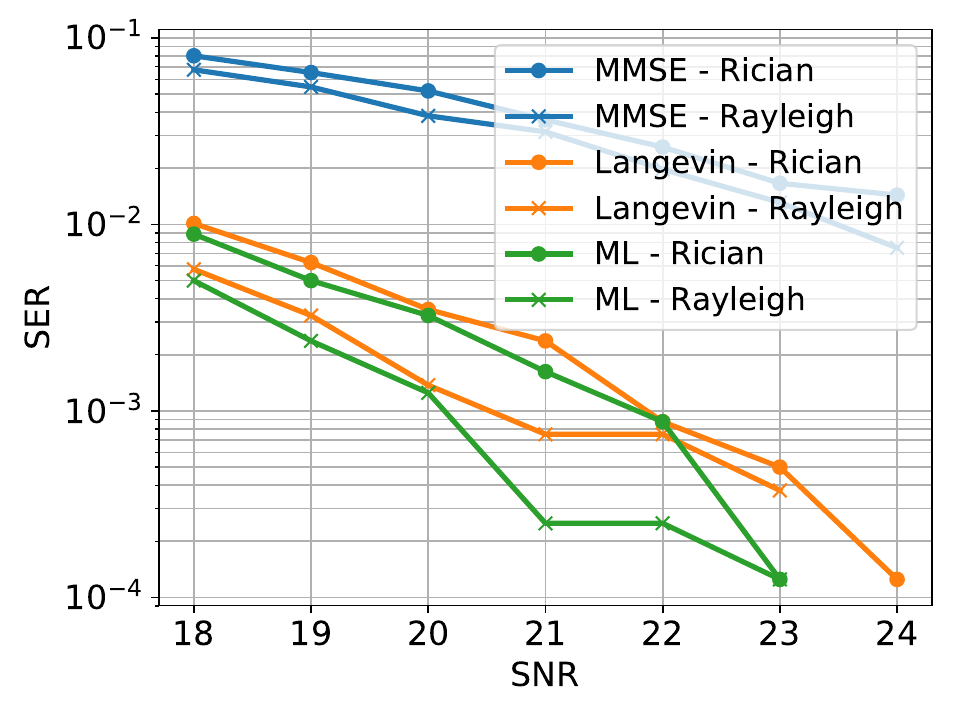}}
    	\vspace{-0.15in}
    	\caption{}
    	\label{fig:SER-nlos}
	\end{subfigure}
	\centering
	\begin{subfigure}{.4\textwidth}
    	\hspace{-0.5cm}{\includegraphics[width=1\textwidth]{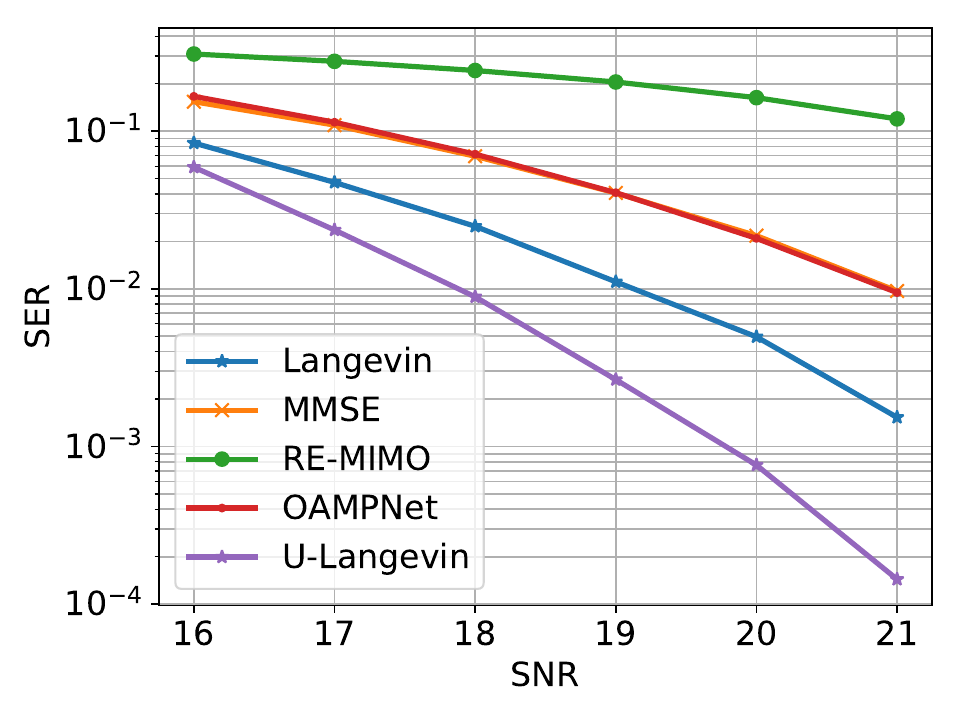}}
    	\vspace{-0.15in}
    	\caption{}
    	\label{fig:SER-unknown_noise}
	\end{subfigure}
	\vspace{-0.02in}
	\caption{ {\small Performance analysis of our proposed method evaluated in a Kronecker correlated channel model as in~\eqref{E:kron}. (a)~SER as a function of SNR for our Langevin method for users transmitting with two 16-QAM and 64-QAM simultaneously. (b)~Comparison of our proposed MIMO detectors with other baseline detector when there is imperfect channel state information. We consider $\text{SNR}=16\text{dB}$ and $\sigma_e \in \{0.002, 0.02, 0.034, 0.06, 0.1\}$. (c)~Comparison of our proposed MIMO detector with the MMSE and ML detectors for channels that belong to the Rician and Rayleigh models. (d)~SER as a function of SNR when there is noise variance uncertainty.}}
	\vspace{-0.1in}
	\label{fig_results_unfold}
\end{figure*}

The result is shown in Fig.~\ref{fig:SER-twomod} for $\ccalX_1 = 16$-QAM and $\ccalX_2 = 64$-QAM.
We compare only with the MMSE, as it is the only low-complexity detector that can handle different modulations simultaneously, and consider two different SNR ranges, one for each modulation. 
The figure reveals that our method outperforms the MMSE detector for both modulations.
However, we observe that the SER performance for the lowest modulation decreases compared to the result in Fig.~\ref{fig:SER-comparison16qamkron}.
On the other hand, the performance for the higher modulation is consistently superior to MMSE for all SNR, remaining a similar behavior to the case of a unique modulation.
Thus, we can conclude that when using different modulation schemes, Langevin consistently outperforms MMSE for all the modulation, although the performance for the users transmitting with lower modulation slightly decreases compared to the case when all users are using the same modulation.

As of last observation, notice that although our experiment was carried out with two modulations, we can seamlessly go to more than two constellations.

%
%
\subsection{Imperfect CSI}
\label{subsec:results_unf}
We evaluate the effect of channel estimation error (imperfect CSI) on the symbol detection performance.
To simulate this scenario, we first generate a batch of channels $\bbH$ from the Kronecker correlated channel model~\eqref{E:kron} and observations $\bby$ using this set of channels; for the learning-based methods, we train using the true channels as well as the observations.
Then, we run all the detectors using a perturbed channel 
\begin{equation}
    \hat{\bbH} = \bbH + \bbE,
\end{equation}
where $\bbE\sim \ccalC\ccalN(0, \sigma_e^2\bbI)$ represents the estimation error.
We consider an $\text{SNR}=16\text{dB}$ for the measurement noise~\eqref{eq:SNRtrue}, and $\sigma_e \in \{0.002, 0.02, 0.034, 0.06, 0.1\}$ for the variance of the estimation error.
The SER as a function of $\sigma_e$ is shown if Fig.~\ref{fig:imperfectCSI}.     
We observe that for low variance, our method still outperforms the other detectors. 
As the error variance grows, all the methods collapse to the same performance.

\subsection{Line-of-sight (LOS) and non-LOS (NLOS) channels}
\label{subsec:nlos_los}

We analyze the SER when we consider a spatially correlated Rician fading model~\cite{NLOSchannel}, which is composed by a deterministic part, known as the LOS component, and a stochastic part that corresponds to the NLOS component. 
Formally, this model is described as follows: given a user $u$, the channel between $u$ and the BS antennas is given by $\bbh_u \sim \ccalC\ccalN(\bar{\bbh}_u, \bbR)$ where $\bar{\bbh}_u$ corresponds to the LOS component and $\bbR$ is the positive semi-definite covariance matrix describing the spatial correlation of the NLOS component.
To get samples from this model, we generate each component separately: first the {NLOS is sampled from $\ccalC\ccalN(\bb0, \bbR)$ and then we add the LOS component $\bar{\bbh}_u$ by choosing some users $u$ randomly.}
Notice that the NLOS component alone corresponds to a Rayleigh fading model.
We consider one BS with $N_r = 32$ and $N_u = 16$, and we use the code in~\cite{NLOSchannel} to generate 500 channels realizations.
The result is show in Fig.~\ref{fig:SER-nlos}.
We compare Langevin with MMSE and ML and we observe that in both settings the proposed Langevin method performs similar to ML.

\subsection{Robustness to noise variance uncertainty}
\label{subsec:results_unf}
The Kronecker correlated channel model~\ref{E:kron} is used and the same simulation setting as in Section~\ref{Ss:baseline}. 
For both OAMPNet and RE-MIMO we trained the model assuming that we know the variance, and then at testing time, we replace the noise variance with an estimation, given by $\hat{\sigma} = \eta \sigma$.
In particular, we consider $\eta = 0.15$ (\si{8\decibel}).
In this case, we generate $M=10$ samples.

The result is shown in Fig.~\ref{fig:SER-unknown_noise}.
We observe that both learning-based methods are highly-dependent on a good estimation of the noise variance.  
On the other hand, the performance of MMSE remains invariant.
Compared to Langevin without learning, the proposed U-Langevin outperforms under this setting of noise uncertainty.
Therefore, the U-Langevin is a more suitable detector in scenarios with noise variance uncertainty as it enhances performance while maintaining the same complexity.
However, in the case of known noise variance, the best option is the Langevin detector without learning as we have closed-form expression and we have shown that it achieves state-of-the-art performance.
%
%


\section{Conclusions and Future Work}
\label{sec:conclusions}
In this paper, we proposed a novel massive MIMO detector based on an annealed version of Langevin dynamics.
The components of the stochastic dynamic -- namely the score of the likelihood and the prior -- were derived in a closed-form solution by leveraging on the annealed process.
This allowed us to naturally incorporate the discrete nature of the constellation elements in the exploration of the search space by the relationship between the score of the prior and the denoiser.
Furthermore, in scenarios where the noise variance is unknown, we relied on the algorithm unrolling framework to parameterize the score of the likelihood.

Revisiting the motivation question in Section~\ref{sec:background}, we demonstrated that combining the two steps in~\eqref{eq:general_iter} and \textit{considering the prior information in the dynamic} entails a massive MIMO detector that outperforms state-of-the-art detection algorithms.
Furthermore, it does not to be trained for a particular channel when the noise is known, making the detector highly flexible to be deployed in real wireless systems.
Through extensive analyses and experiments, we have demonstrated that the annealed Langevin detector yields a solution that is: i) Accurate, as it outperforms state-of-the-art detectors using different channel models; ii) Flexible, as it can handle different modulations simultaneously; iii) Easy to be deployed, as it was shown in the real channel experiment.

The main limitation of the proposed detector is the running time and complexity, in particular for fast-fading channels. 
A possible research direction to improve the running time is to explore alternative stochastic diffusion processes that work in non-asymptotic regimes.
This will allow to reduce the number of iterations while still sampling from the posterior distribution considering the discrete nature of the symbols.
From the theoretical point of view, future work includes leveraging the rich Langevin theory to derive non-asymptotic convergence theoretical guarantees.
Finally, from the experimental side, testing the unfolded detector with non-Gaussian noise using the unfolded detector is left as future work.

\vspace{-1mm}






\bibliographystyle{IEEEbib}
\bibliography{citations}

\end{document}